\documentclass[12pt,preprint]{aastex}

\usepackage[dvips]{color}

\slugcomment{Draft \today}

\newcommand{\gI}{$g\!-\!I$}
\shorttitle{Globular clusters in Four Poststarburst Galaxies}
\shortauthors{Maybhate et al.}

\begin{document}

\title{Globular Cluster Populations in Four Early-Type Poststarburst Galaxies
  \altaffilmark{1}}

\author{Aparna Maybhate\altaffilmark{2}, Paul Goudfrooij\altaffilmark{2},
  Fran\c cois Schweizer\altaffilmark{3}, Thomas H. Puzia\altaffilmark{4},
  David Carter\altaffilmark{5}} 

\altaffiltext{1}{Based on observations with the NASA/ESA {\it Hubble
    Space Telescope}, obtained at the Space Telescope Science
  Institute, which is operated by the Association of Universities for
  Research in Astronomy, Inc., under NASA contract NAS5-26555} 
\altaffiltext{2}{Space Telescope Science Institute, 3700 San Martin
  Drive, Baltimore, MD 21218; maybhate@stsci.edu, goudfroo@stsci.edu} 
\altaffiltext{3}{Carnegie Observatories, 813 Santa Barbara Street,
  Pasadena, CA 91101; schweizer@ociw.edu} 
\altaffiltext{4}{Plaskett Fellow, Herzberg Institute of Astrophysics, 5071 West Saanich
  Road, Victoria, BC V9E 2E7, Canada; puziat@nrc.ca} 
\altaffiltext{5}{Astrophysics Research Institute, Liverpool John
  Moores University, 12 Quays House, Egerton Wharf, Birkenhead, CH41 1
  LD, United Kingdom; dxc@astro.livjm.ac.uk}

\begin{abstract}
We present a study of the globular cluster systems of four early-type
poststarburst galaxies using deep {\it g}- and {\it I}-band images from the
{\it ACS} camera aboard the {\it Hubble Space Telescope (HST)}.
All the galaxies feature shells distributed around their main bodies and are
thus likely merger remnants. 
The color distribution of the globular clusters in all four galaxies
shows a broad peak centered on $g-I \approx 1.4$, while PGC 6240 and
PGC 42871 show a significant number of globular clusters with
$g\!-\!I \approx 1.0$.
The latter globular clusters are interpreted as being of age $\sim$\,500
Myr and likely having been formed in the merger. 
The color of the redder peak is consistent with that expected for
an old metal-poor population that is very commonly found around
normal galaxies. However, all galaxies except PGC 10922 contain
several globular clusters that are significantly brighter than the
maximum luminosity
expected of a single old metal-poor population.
To test for multiple-age populations of overlapping \gI\
color, we model the luminosity functions of the globular clusters
as composites of an old metal-poor subpopulation with a range of plausible
specific frequencies and an intermediate-age subpopulation of solar
metallicity. We find that three of the four sample galaxies show evidence for
the presence of an intermediate-age ($\sim$\,1 Gyr) globular cluster
population, in addition to the old metal-poor GC population seen in normal
early-type galaxies. None of the galaxies show a significant population of
clusters consistent with an old, metal-rich red cluster population that is
typically seen in early-type galaxies.

The presence of a substantial number of intermediate-age clusters and
the absence of old, metal-rich clusters indicate that the progenitor
galaxies which formed the resulting shell galaxy were gas rich
and did not host significant bulges. Late-type spirals seem to be the most
plausible progenitors. These results lend credence to the
``merger scenario'' in which the red, metal-rich GCs
observed in normal ellipticals are formed during a dissipative merger event 
that also forms the elliptical itself.

\end{abstract}


\section{Introduction}

Mergers seem to have played a major role in determining the shapes and
dynamics of elliptical galaxies. A few galactic mergers still occur and
offer valuable clues to past evolutionary processes. Young globular
clusters formed during mergers hold strong promise for age-dating such
events, besides helping shed light on the cluster-formation process itself.
Globular clusters are very useful probes of the dynamical and chemical
assembly history of galaxies. Many globular cluster (GC) systems in normal
giant elliptical galaxies show a bimodal color distribution, indicating
the occurrence of a second event/mechanism of cluster formation.
The ``merger'' model suggests that metal-rich (``red'') clusters are formed
during major mergers of gas-rich galaxies \citep{sch87,az92}.
Globular cluster systems in young merger remnants such as NGC 7252
\citep[e.g.,][]{miller97} reveal a bimodal color distribution featuring not
only the well-known ``universal'' halo population of old metal-poor GCs, but
also hundreds of second-generation GCs that are young, metal-rich, and have
most probably formed from the metal-enriched gas associated with the
progenitor spirals. The many structural properties that merger remnants such
as NGC 1275, NGC 3921, and NGC 7252 share with normal ellipticals suggest not
only that these merger remnants are proto-ellipticals
\citep[e.g.,][]{schweizer98} but also that many normal ellipticals with
bimodal color distributions may have formed their metal-rich GCs in a similar
manner. If this is indeed the case, it should be possible to find ellipticals
with second-generation GC systems of intermediate age ($\sim$ 1--4
Gyr). Evidence for the existence of intermediate-age GCs in ellipticals has
been recently found for a few galaxies like NGC 1316
\citep{goudfroo01a,goudfroo01b,goudfroo04}, NGC 3610
\citep{whitmore97,whitmore02,goudfroo07}, NGC 1700 \citep{whitmore97,brown00},
NGC 4365 \citep{puzia02,larsen03,kundu05}, and NGC 5128 \citep{minniti96,peng04}.  

This paper aims at searching for intermediate-age clusters in a sample
of early-type galaxies which show evidence for having experienced a merger
and a starburst in the recent past (during which GCs may well have
formed). To select the sample, we used the \citet{malin} catalog. This catalog
consists of southern elliptical galaxies that feature shells
and/or ripples around their main body and in their outer envelopes. Shells
are a common result of galaxy mergers \citep{quinn84,dupraz86,hq88,hq89}
or interactions \citep{tw90,thomson91} and are expected to have a
lifetime of only $\sim 10^9$ years. Approximately 10\% of these 
early-type shell galaxies have colors, absolute magnitudes, and spectra
characteristic of ``poststarburst'' 
galaxies \citep[featuring strong Balmer lines][]{carter88}.   

A poststarburst spectrum is characterized by the presence of
strong Balmer absorption lines (indicative of A-type
stars), but without strong [O\,{\sc ii}] or H$\alpha$ emission lines
\citep{dressler83}. The existence of strong Balmer absorption lines
indicates that these galaxies have experienced star formation in the
past $\sim$ 0.1 -- 1 Gyr while the absence of emission lines indicates
they are no longer forming stars.  Most such galaxies are classified as
poststarburst based on the nature of the spectrum only in their central
region.  Often no information is available on the extent of the poststarburst
event or the amount of {\it global} star formation.
It is interesting to investigate the global properties of the poststarburst class of
galaxies and explore the connection (if any) between properties of the host galaxy
and its star formation history. We will address these issues through a
study of the GC systems of these galaxies since they provide
a good handle on the impact of major star formation episodes in a galaxy.
If numerous intermediate-age GCs with 
ages consistent with those of the poststarburst population in the
nucleus were to be found, we could be fairly confident that the A
stars were formed in a relatively vigorous star formation
event, likely associated with a recent dissipative galaxy merger. Conversely,
an absence of such GCs would indicate that no strong star formation was
associated with the event that caused star formation to end $\sim$\,0.1 to
1 Gyr ago. The latter would be consistent with a scenario where gas is
removed from the galaxy, e.g.\ by weak ram pressure stripping.

The nearest four galaxies from \citet{carter88} which showed a poststarburst
spectrum were selected for this study. The basic properties of these galaxies
are given in Table~\ref{tbl-1}.  A detailed study of one of the galaxies,
namely PGC 6240 (AM 0139$-$655) has been presented in an earlier paper
\citep[][hereafter Paper I]{aparna}. 
The current paper presents a study of the properties of the GC systems as well as the
diffuse light of all four early-type poststarburst
galaxies.  It discusses the implications of these properties in the context
of the star formation history of these galaxies, 
and of the assembly history of early-type galaxies in general.

\section{Observations and Data Reduction}

\begin{figure}
\plotone{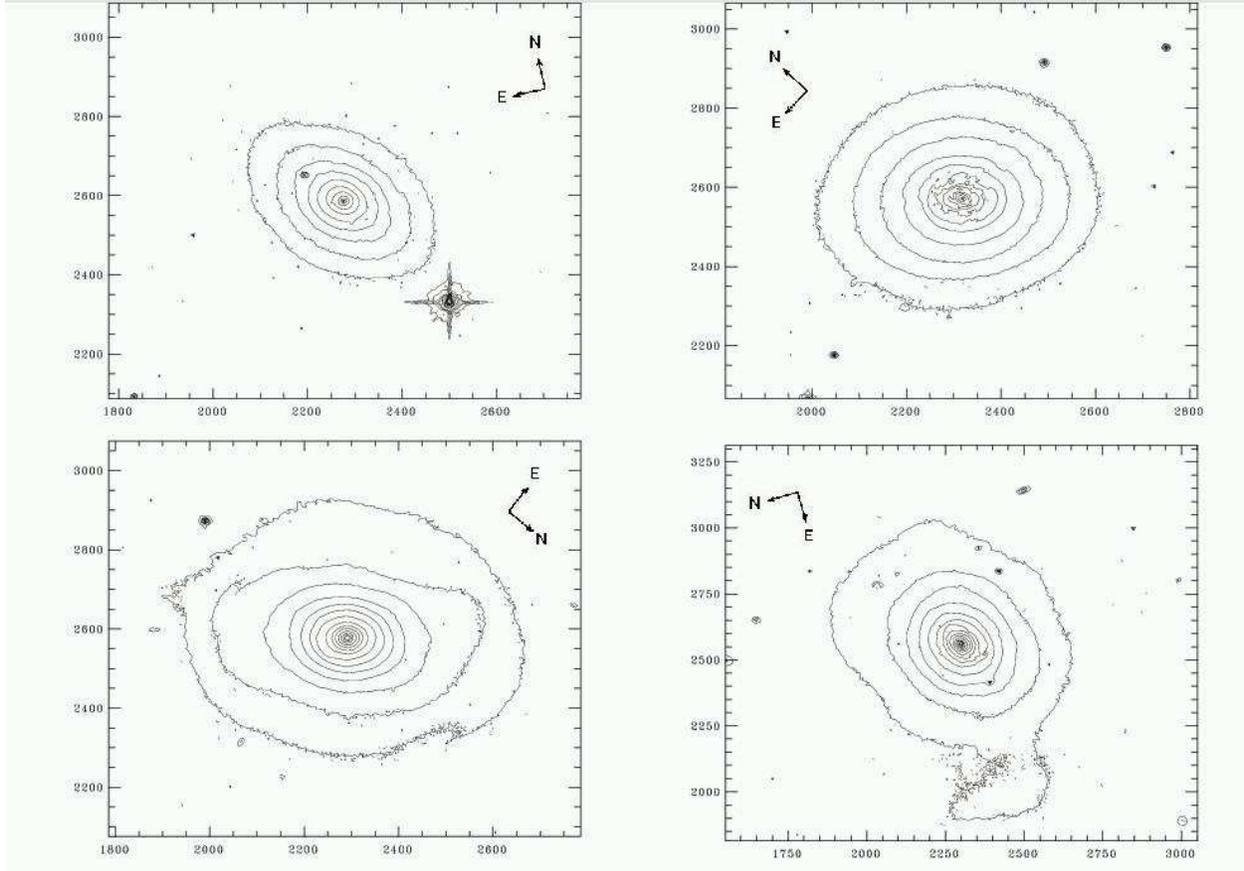}
\caption{Contour plots of the sample galaxies.  Contours represent $I$-band
  surface-brightness values from 13.5 mag arcsec$^{-2}$ to 21.5 mag arcsec
  $^{-2}$ for PGC 6240 ({\it bottom right}) and 13.5 mag arcsec$^{-2}$ to
  20.5 mag arcsec$^{-2}$ for the remaining galaxies ({\it top left:} PGC 6510,
  {\it top right:} PGC 10922, and {\it bottom left:} PGC 42871).
  In spite of several common properties
  like morphological type, poststarburst spectrum, and presence of shells,
  the isophotes show a substantial variety among the four galaxies. \label{contour}}
\end{figure}

The four galaxies were observed as part of HST
General Observer program 10227 (PI: Goudfrooij), using the Wide Field Channel
(WFC) of ACS with the filters F475W and F814W.
The observations consisted of several long exposures plus a few short ones
to guarantee getting an unsaturated galaxy center. The total exposure times
were 7874s, 9350s, 8369s, and 21440s in F475W and 2712s, 3970s, 5908s, and 7300s
in F814W for PGC 6510, PGC 10922, PGC 42871, and PGC 6240, respectively.
The data were processed with the ACS on-the-fly pipeline, which
included dark and bias subtraction and flat fielding.  Individual flat-fielded
images were carefully
checked for satellite trails and saturated pixels in the central region
of each galaxy. These pixels were flagged and masked out.
The individual images of a galaxy
in each band were co-added using the PyRAF\footnote{PyRAF is a product
  of the Space Telescope Science Institute, which is operated by AURA
  for NASA} task MULTIDRIZZLE \citep{kok}. This resulted in images
cleaned of cosmic rays and corrected for geometric distortion.  A combination
of the short and long-exposure images enabled us to get good resultant
multidrizzled images with unsaturated centers. The isophotal contours of the four galaxies
are shown in Fig.~\ref{contour}. 
It is obvious from the figure that the sample galaxies show significant
differences in global morphology, even though they all share the properties of a
poststarburst spectrum and shell structure. PGC 6510 appears to be a disky
elliptical, PGC 10922 shows rounder isophotes but has a distorted center (due
to dust absorption), while the isophotes of both PGC 6240 and PGC 42871 appear
irregular in the outer regions.

\section{Cluster Candidate Selection}

To select cluster candidates, 
the F475W and F814W images obtained from MULTIDRIZZLE were first
added together to get a high signal-to-noise {\it g+I\/} image for each galaxy. Elliptical
isophotes were then fit to this coadded image using the task ELLIPSE within
STSDAS\footnote{STSDAS is a product of the Space Telescope Science Institute,
  which is operated by AURA for NASA}, and allowing the center, 
ellipticity, and position angle of the isophotes to vary.
This yielded a smooth model of each galaxy's count distribution.
The search for sources
was performed on an image created by dividing the {\it g+I\/} image
by the square root of the model image.  This ensures uniform shot noise
characteristics over the whole image. The detection threshold for selecting
sources was set at 4\/$\sigma$ above the background. 
Typical half-light radii of globular clusters fall into
the range 1 -- 20 pc \citep{kundu,vanden,jordan}.  At the distance of our
sample galaxies, the spatial scale is 14\,--\,25 pc per ACS$/$WFC pixel.
Thus, we expect the globular clusters to appear as nearly unresolved point
sources. 
There are two advantages in using the {\it g+I\/} images, rather than the
individual $g$- or $I$-band images, for cluster-candidate selection.
Firstly, the coadded images reach a greater depth than the individual images,
and secondly, the photometric zero-point of a {\it g+I\/} image is
significantly less color dependent than that of the individual images.
A detailed illustration and 
discussion of these advantages can be found in \citet{goudfroo07}.

To perform cluster photometry, a
smooth elliptical model was constructed for each galaxy in each
filter in a manner similar to the one described above. This model was
subtracted from the corresponding drizzled image to get a residual image. The
cluster candidates are clearly visible once the underlying galaxy light has
been subtracted.  Aperture photometry was then performed through an aperture
of 3 pixel radius for all the detected sources. The photometry lists thus
obtained for
both filters were then matched for further analysis. Aperture corrections
from 3 pixel to 10 pixel radius were determined using a few bright
point sources in
each band.  The corrections from 10 pixel radius to infinity were taken from
\citet{siri}.  Finally, the F475W and F814W magnitudes were converted from
the instrumental system (STMAG) to SDSS {\it g} and Cousins {\it I} magnitudes
in the VEGA system
via the SYNPHOT package in STSDAS; for details, see Paper I. The {\it g} and
{\it I} magnitudes and the $g-I$ color were corrected for Milky Way foreground
reddening using  $A_V$ given in Table~\ref{tbl-1} and the relations 
$A_g = 1.16 \times A_V$ and  $E(g-I) = 0.59 \times A_V$.

After discarding clusters with photometric errors $>$\,0.4 mag in each
band, the first selection criterion we applied considered the target's
color index $g-I$.  Using the population-synthesis models GALEV of
\citet{galev}, we determined that
the range  $0.0 \leq g-I \leq 2.2$ includes all model clusters 
with ages from $10^{7}$ to $10^{10}$ years and metallicities $Z$ from
0.0004 to 0.05 (where $Z=0.02$ corresponds to solar metallicity), as shown in
Fig.~\ref{ssp}.  The GALEV models are calibrated to the ACS filters and
were hence chosen to avoid introducing systematic errors while converting to
a standard system. 
However, the differences between model predictions of GALEV and
(e.g.) \citet{bc03} in the age range 0.5\,--\,14 Gyr stay within 0.05
mag in $V\!-\!I$. We also verified that the results obtained using
the \citet{maraston} models are consistent with those from GALEV as shown
in Fig.\ 2.
An extensive comparison between the various Simple Stellar Population (SSP)
models can be found in \citet{pessev}.
Sources with colors outside the range $0.0 \leq g-I \leq 2.2$ were judged
unlikely to be real star clusters and were discarded.
The resultant source list contains cluster candidates, stars, and background
galaxies.  To further constrain the selection and discard extended
background galaxies, we determined the FWHM of all objects in the
color-selected lists from the coadded image, using the IRAF task IMEXAM, and
retained objects with $1.5 <$ FWHM $< 3.0$ pixels. An additional
compactness criterion was applied to only retain sources that had a
difference in magnitudes between apertures of 2 and 5 pixel radius
of 0.2 to 0.6 mag.  Our final list of GC candidates contains the sources
that satisfy all the above criteria.  

\begin{figure}
\plotone{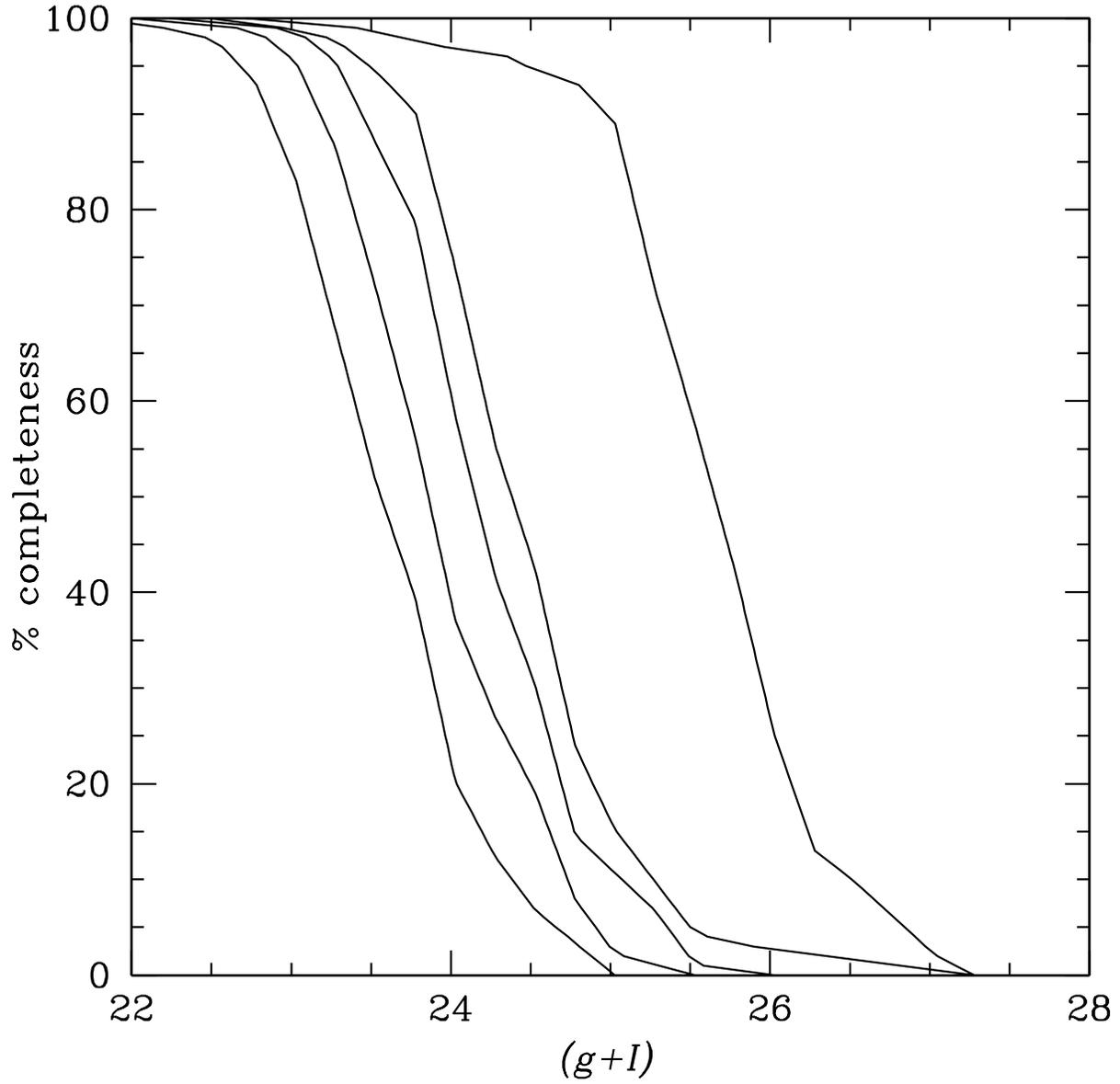}
\caption{A set of typical completeness curves determined for the combined
{\it g+I\/} image for five different values of the background. The curves
shown are for PGC 6510 and represent from left to right: 1000, 530, 276,
150, 80 counts per pixel for an effective {
\it g+I} exposure time of 705 s.\label{ccurves}}
\end{figure}

Completeness tests were done for each galaxy on the image used to detect
cluster candidates
(i.e., the {\it g+I\/} image).  Artificial clusters were added to the image in
batches of 100 for five different background levels and in 0.25 mag intervals.
Using the
background levels in the parts of the image farthest away from, and
nearest to, the galaxy center where clusters are still detected as the
two extremes, the five background levels were determined by dividing this
interval into five logarithmically equal intervals.  The radial-intensity
profile of the artificial clusters was determined by fitting PSFs to the
real clusters in
the {\it g+I\/} image.  A smooth elliptical model of each host galaxy was
obtained, and the {\it g+I\/} image was divided by the square root of
this model image.  Source detection was then performed on this image in
the same manner as done previously for the actual sources. 
Other criteria (permissible error in photometric magnitude, FWHM, and
compactness) were also applied in exactly the same manner as before.
A typical sample of completeness curves, obtained for PGC 6510
is shown in Fig.~\ref{ccurves}.

\subsection{Selecting ``Genuine'' versus ``Background'' sources}

\begin{figure}
\plotone{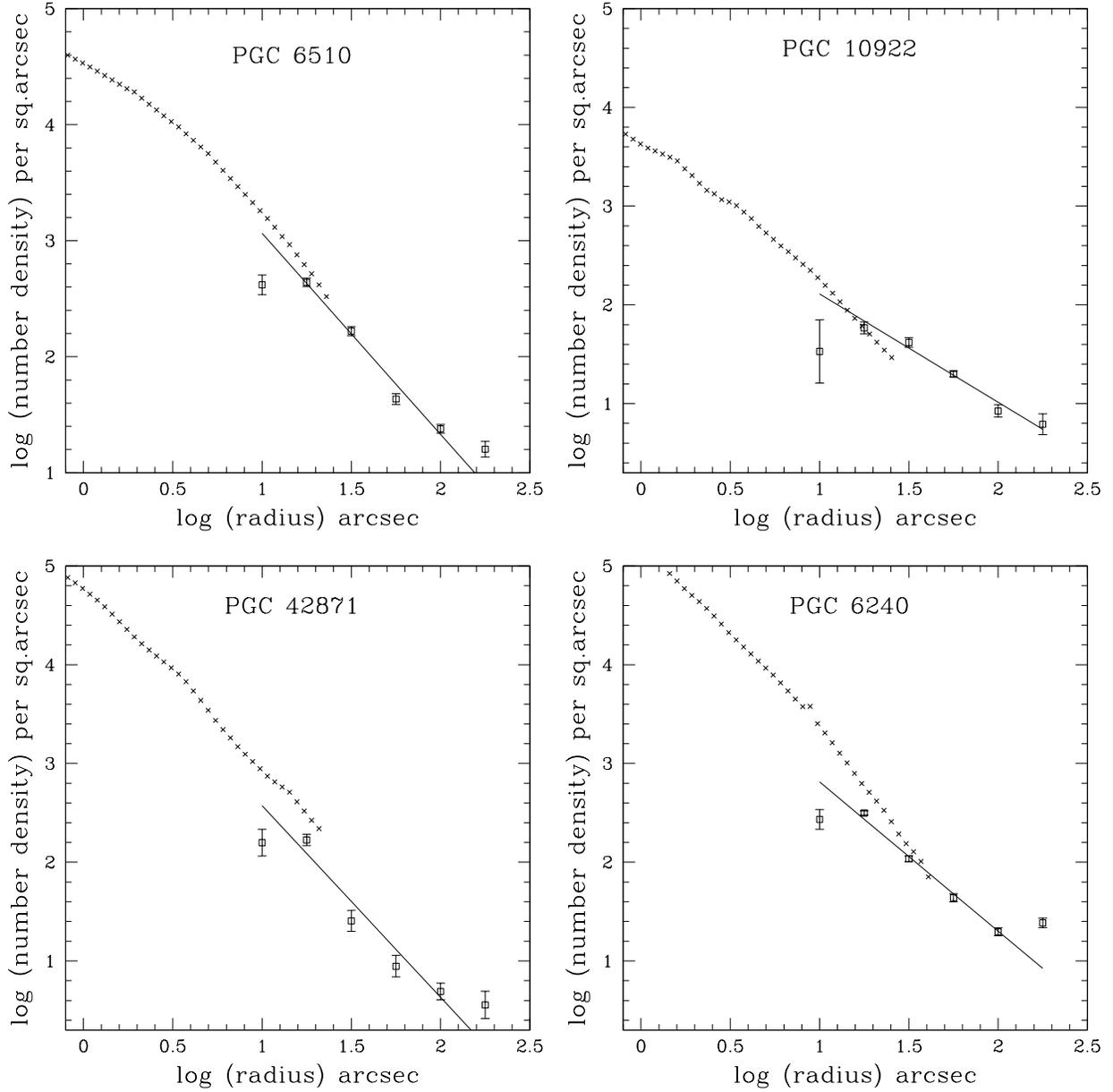}
\caption{The surface number density of the GC candidates is compared to the
  surface brightness profile of the underlying galaxy. The crosses denote the
  surface brightness of the underlying galaxy light with an arbitrary
  zeropoint. The
  squares represent the logarithmic number density of GC candidates per
  square arcsec. The solid lines represent power-law fits to the GC surface
  number density in the outer regions. \label{magdensity}}
\end{figure}

To determine the radial extent of the
cluster candidates that are physically associated with
the target galaxy versus compact background galaxies that have similar
colors and apparent magnitudes, we examined the surface number density of the GC
candidates as a function of galactocentric radius. This was achieved
by dividing each galaxy image into annular rings centered on the galaxy center and
computing the number of sources per unit area in each ring.  The bright
background due to the presence of the galaxy would prevent the detection of
faint sources in the inner regions, while sources of the same magnitude would
be easily detected in the outer regions of low background. 
To account for this effect, we applied a magnitude cut-off to the sources
that were considered for the surface density measurements.  This
cut-off was chosen to be at the magnitude associated with a
completeness value of 80\% at a galactocentric radius of 6\arcsec. 
Only sources outside that radius were taken into account.  
Figure~\ref{magdensity} shows the results.  We find that the
surface-density profile flattens off beyond a certain radius for each
galaxy. This radius was taken to be the limiting radius for the
sources associated with that galaxy and was found to be 60{\arcsec},
94{\arcsec}, 104{\arcsec}, and 78\arcsec\ for PGC 6510, PGC 10922,
PGC 42871, and PGC 6240, respectively.  All sources detected beyond this
radius were considered background sources. 

The radial GC surface-number-density profiles are in most cases 
similar to the surface brightness profile of the parent galaxies, which is
consistent with the situation for the metal-rich subpopulation of GCs
in normal early-type galaxies
\citep{harris79,harris86,az98,goudfroo07}. However, the GCs in PGC 10922 seem
to have a more extended distribution, similar to that seen for metal-poor GCs
in normal early-type galaxies. Quantitatively, simple power-law fits to the
outer data points yield exponents ranging from $-1.09$ to $-1.94$, whereas
surface-number-density
profiles for GCs in normal elliptical galaxies show power-law exponents in the
range between $-1$ and $-2$ \citep{az98,puzia04}. It is also seen that in all
cases, the innermost GC bin has a lower surface number density than
that represented by the power-law fits. This could be due to the faster
destruction of GCs in the inner regions of these galaxies due to effects
such as bulge shocking
\citep[e.g.,][]{gnedin98,puzia04}. Table~\ref{tbl-2} gives the 
limiting magnitude, power-law exponent, and radial range considered for the
fit in each case.  A list of the thirty brightest clusters in the $I$-band
in PGC 6510, PGC 10922, and PGC 42871 (with their positions, $I$-band
magnitudes and distance from the galaxy center) is given in Table~\ref{tbl-3}.
Paper I already contains a list of the brightest clusters in PGC 6240 and
it is not repeated here.

\section{Globular Cluster Color Distributions}

\begin{figure}
\plotone{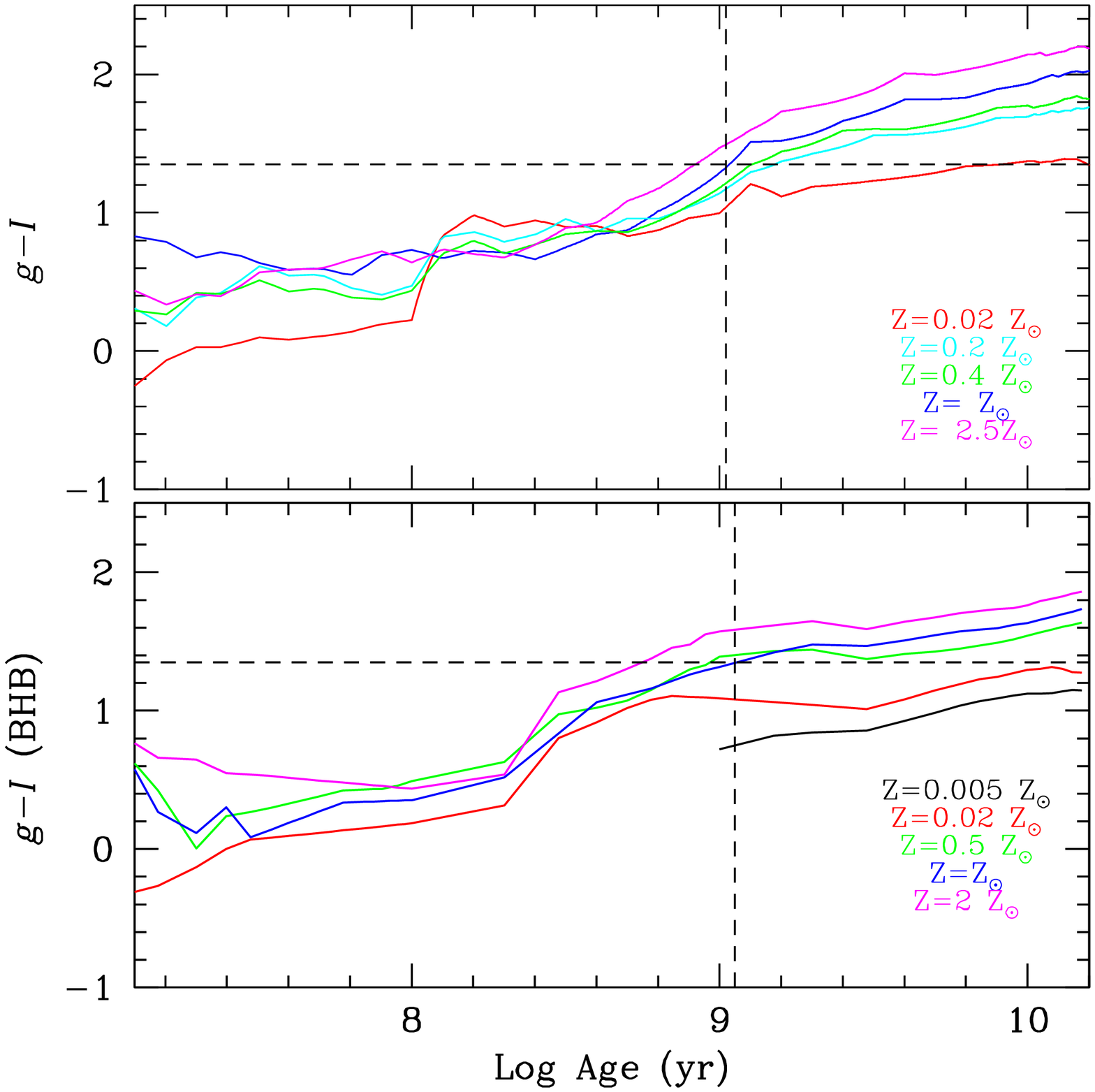}
\caption{{\it Top panel}: Time evolution of the \gI\ color index using GALEV SSP models
  \citep{galev}. Model curves are plotted for a \citet{sal} IMF and
  metallicities as indicated in the figure. \gI\ = 1.35 is consistent with
  both the universal metal-poor population (Z $\sim$ 0.02 Z$_\odot$) as well
  as a 1 Gyr population with solar metallicity.  {\it Bottom panel}: Time evolution of
the \gI\ color index for the blue horizontal branch using Maraston models \citep{maraston}. The ages
obtained for \gI\ = 1.35 is consistent with that
obtained using the GALEV models. \label{ssp}}
\end{figure}

\begin{figure}
\plotone{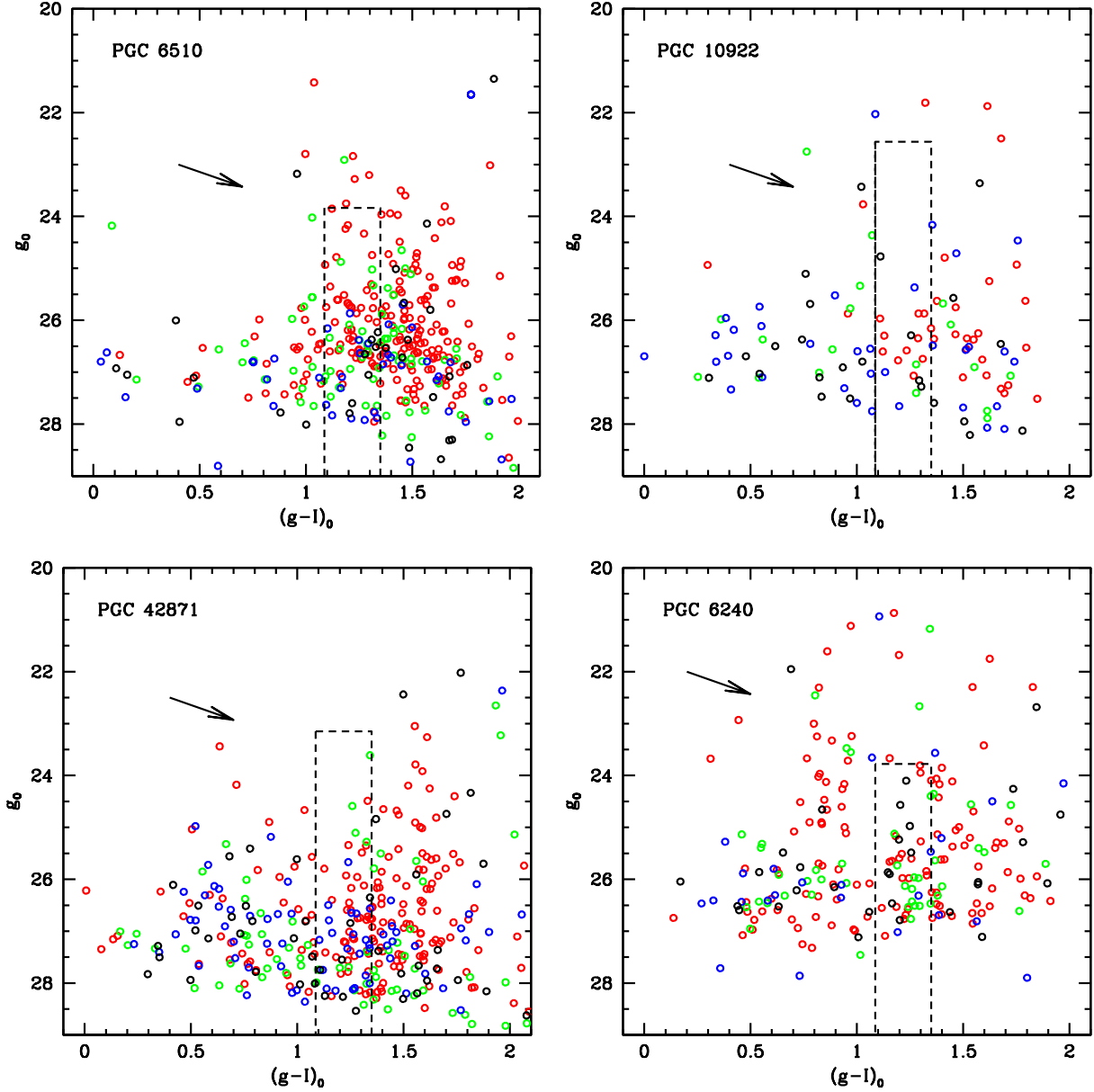}
\caption{$g$ versus \gI\ color--magnitude diagrams of the GC
  candidates in each galaxy within the limiting radius given in
  Sect. 3.1. The red, green, blue, and black circles represent
  GC candidates within equal areas of increasing galactocentric radius with the
red ones found at the smallest radii. The rectangle enclosed by dashed lines
 represents the
  magnitude and color range expected for old metal-poor GCs, and the arrow is
  the
 vector for dimming by age from 1 Gyr to 1.6 Gyr as determined from the GALEV models.\label{cmd}}
\end{figure}

Color--magnitude diagrams for the clusters associated with each galaxy within
the limiting galactocentric radius as found in the previous section are shown
for equal areas of increasing galactocentric distance in Fig.~\ref{cmd}. 
Note that the magnitudes and colors have been corrected for foreground reddening
as described in Sect.\ 3.
For comparison, we indicate the area which would be populated by old
metal-poor GCs similar to those in the halo
of our Galaxy, scaled to the distance of each galaxy. 
Typically a large number of GCs with $g\!-\!I \approx 1.4$ to 1.5 is seen
  in each galaxy.
This color index is similar to, though
slightly larger than,
the mean index expected for a 
population of old metal-poor GCs like those in the  halo of our
Galaxy (e.g., a mean age of 14 Gyr and a mean [Fe/H] of $-$1.5).  Note,
however, that this color is also consistent with solar-metallicity
clusters of age $\sim$\,1 Gyr as indicated by
SSP synthesis models (cf.\ Fig.~\ref{ssp}), the presence of which
might be expected in a poststarburst galaxy. 
Intermediate-age GCs in established merger remnants feature a radial
surface number density distribution that is significantly more
centrally concentrated than that of old 
metal-poor GCs \citep[e.g.,][]{whitmore97,goudfroo07}.  
In an attempt to discriminate between the two plausible interpretations of the
nature of the GCs with $g\!-\!I \approx 1.4$, we therefore 
divide up the clusters into four equal-area bins of increasing
galactocentric radii (see different symbols in Fig.~\ref{cmd}).
All color distributions appear broader than the halo
GC distributions in our Galaxy as outlined by the rectangles.
The innermost GCs in most of the galaxies have mean colors that
are somewhat redder ($g\!-\!I \approx 1.5$) than those of old metal-poor GCs.  
While this could in principle be a result of reddening by dust in the
inner regions that shifts the clusters to fainter magnitudes and
redder colors, there are 
very few clusters found in the inner dusty regions. Alternatively, it could 
indicate the presence of a second-generation population of clusters
with ages slightly larger than 1.0 Gyr. 
A comparison of the dimming vector for a 1.0 Gyr old population aging to
1.6 Gyr (shown in 
Fig.~\ref{cmd}) with the distribution of the inner clusters in the CMD
does suggest the possibility of these clusters being slightly
older than 1.0 Gyr. Apart from that, there is also a population of (inner)
clusters with colors significantly bluer than old metal-poor GCs
in PGC 6240. Thus, it appears likely that the GC systems of these poststarburst
galaxies are made up of more than a single-age population. The question of
how to disentangle the different populations in a quantitative way
will be discussed  further in the next Section.    

\begin{figure}
\plotone{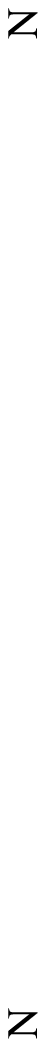}
\caption{GC color distributions for each of the sample galaxies. The open
  histogram in each panel represents the observed color distribution
corrected for foreground reddening, while the
  hatched histogram represents the background-subtracted distribution. The
  distribution of the background sources is shown in the lower panel for each
  case. The solid black line is the probability-density estimate of the
  uncorrected distribution. The solid red line is a non-parametric
  probability-density estimate using an adaptive Epanechnikov kernel of the
  background-corrected GC color distribution. The dashed red lines mark the
  bootstrapped 90$\%$ confidence limits. \label{color}}
\end{figure}

The color distribution of the GCs was analyzed by first computing the
non-parametric
Epanechnikov-kernel probability-density function \citep{silverman86}
of all objects within the limiting radius.
The probability-density function of all objects outside this radius was
then computed similarly
to provide an estimate of the contamination by compact background galaxies.
Finally, the background density estimate was scaled to that of the cluster
candidates by area, and the true color
distribution of the GCs was derived by statistical subtraction.
Figure~\ref{color} shows the color distributions of the sources within the
limiting radius, the background sources, and the inner sources after
statistical background subtraction.
The final color distribution appears mostly unimodal in
PGC 6510, while that of PGC 42871 shows the possible presence of a second,
bluer population similar to that found earlier for PGC 6240 \citep{aparna}.
According to the SSP models of \citet{galev} shown in Fig.~\ref{ssp},
these can be explained by a metal-rich population with ages of a few
hundred Myr.
In case of PGC 10922, the color distribution shows some indications
of clusters with colors $\approx$ 1.0 and $\approx$ 0.0. However, as they
are very few in number, it is difficult to draw any conclusions about
them.
It is also interesting to note that we do not detect any significant
population of old, metal-rich ($-1.0 \la \mbox{[Fe/H]} \la 0.0$) GCs in any
of the sample galaxies (they should show up at
$1.8 \la g-I \la 2.0$), whereas such a population {\it is\/} generally present 
in normal early-type galaxies \citep[e.g.,][]{peng06}. This is discussed
further in Section~\ref{s:HostConnection}. 

\section{Luminosity Functions}

\begin{figure}
\plotone{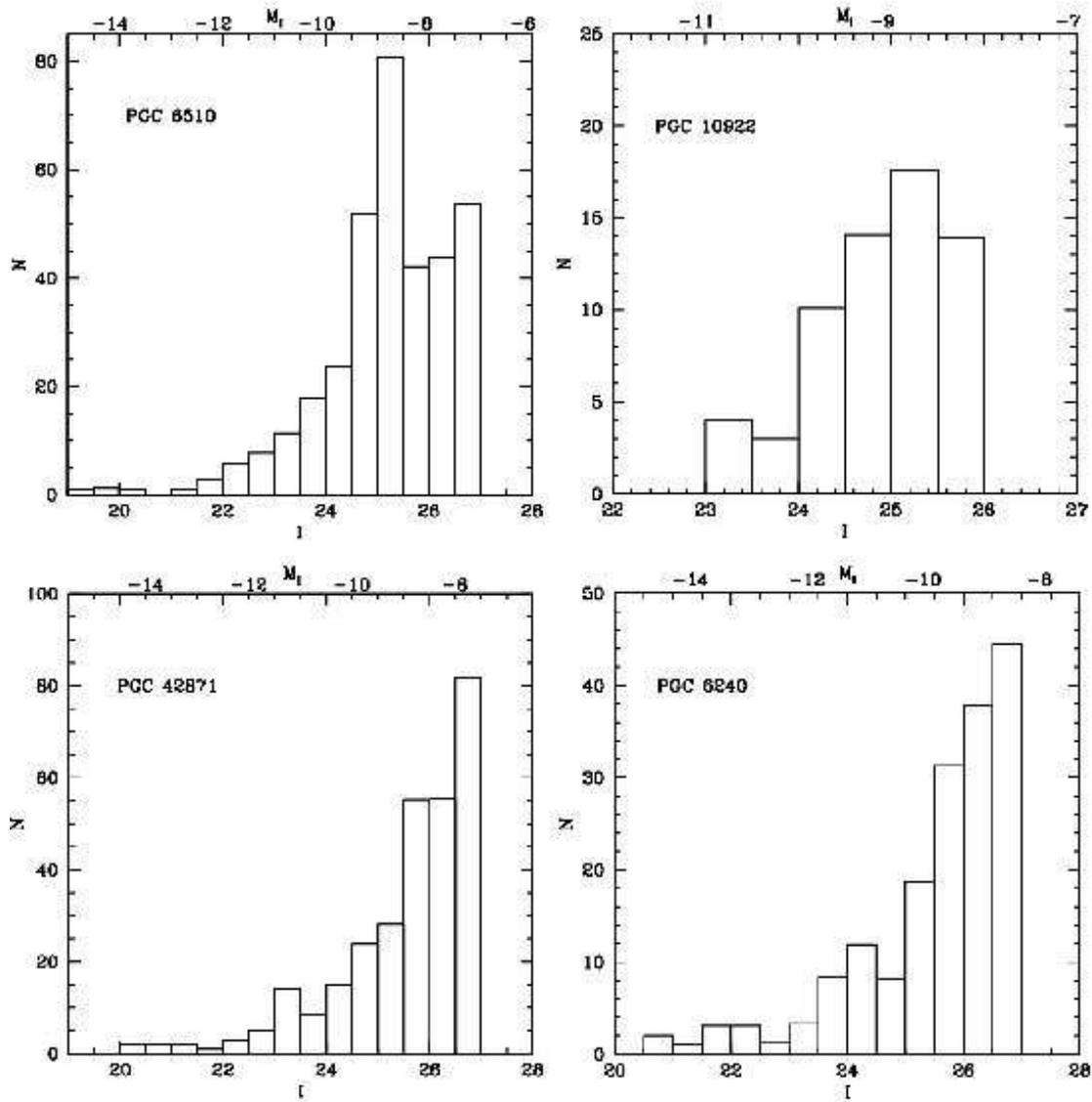}
\caption{Luminosity functions of the GCs in the $I$-band corrected for
  background contamination and completeness. The $I$-band magnitudes have
been corrected for foreground reddening. Note the presence of several
  luminous clusters brighter than $M_I = -11.0$ in PGC 6510, PGC 42871, and
  PGC 6240. \label{lf}}
\end{figure}

To derive cluster luminosity functions (LFs), we
used the completeness curves described in Sect.\ 3 to assign a completeness
value to each cluster.  This value was computed via bilinear
interpolation in cluster magnitude and background value.  Clusters with
completeness values less than $25\%$ were excluded from being counted.
The remaining clusters were divided into two groups based on their distance
from the center of the galaxy.  Clusters within the limiting radius
(determined in Sect.\ 3) were designated as actual GC candidates and those
outside this
radius as likely background contaminants. The LFs of the GC candidates were
then corrected for background contamination using the scaled LFs of the
background. The final corrected LFs are shown in Fig.~\ref{lf}. 

If the LFs were made up entirely of clusters belonging to an old metal-poor
population similar to that found in our Galaxy, and hence had a gaussian form
with a turn-over at $M_I = -8.1$ and $\sigma = 1.2$ \citep{harris96,
  barmby01}, we would not expect to see any clusters brighter than
$M_I \approx -11.0$.
However, except for PGC 10922, the galaxies in our sample show
clusters significantly brighter than this value. 
While the GC color distribution alone may show no obvious indication of the
presence of more than one subpopulation, the fact that overluminous
GCs are seen in the LF which are unlikely to belong to an 
old population suggests that the observed
LF is due to the superposition of an old population and a younger population
with mean colors that are similar to one another.

For example, the LF of PGC 42871 shows an excess of luminous clusters
at magnitudes brighter than $M_I = -11.0$. 
Inspection of the CMD (Fig.~\ref{cmd}) shows that these luminous clusters
predominantly have \gI\ colors redder than 1.5 mag and, hence, are not
associated with the bluer peak seen in Fig.~\ref{color}. 
SSP models show that between the ages of 1 and 2 Gyr, the colors of
metal-rich 
clusters of solar metallicity are indistinguishable from those of typical old
metal-poor clusters. However, these intermediate-age clusters are brighter
than the old clusters by $\sim$1.5 mag for a given mass
\citep{whitmore97,goudfroo01b}.  
Using the SSP models of \citet{galev} shown in Fig.~\ref{ssp}, we find that
the colors of these luminous clusters are consistent with them being
intermediate-age clusters (1\,--\,1.5 Gyr) of solar metallicity.

Our assumption of solar metallicity
for these second-generation GCs is based on (a) the expectation of finding
high metallicity GCs formed from gas in evolved spirals which are the
likely precursor galaxies, and (b) the finding that intermediate-age
GCs in other merger remnants have metallicities consistent with solar
\citep[e.g.,][]{ss98,goudfroo01a,schweizer04}. 

\subsection{Modeling the Composite LF} \label{s:compositeLF}

As the above example shows, the color distribution of clusters in
poststarburst galaxies of age $\sim$\,0.5\,--\,2 Gyr is not expected to
show the signature of two subpopulations prominently because the younger,
metal-rich population will have a mean color similar to that of the
old metal-poor population. 
In that case, one way to test for the presence of multiple populations
and estimate their properties is to study the LF. 
In Paper I, we attempted to model the cluster LF of PGC 6240 as a combination of an
old metal-poor population and an intermediate-age population. The old
population was estimated by fitting template gaussian LF components.
These were then
subtracted from the observed LF, and the residual LF was used to
estimate the contribution of the intermediate-age population by means of
least-squares fits.   
Here we use the same method to investigate the LFs of clusters in
PGC 6510, PGC 10922, and PGC
42871 and to explore the possibility of multiple cluster populations
of different ages but similar color.  For PGC 10922 and PGC 42871, we
consider only clusters with \gI\ $> 1.15$ to avoid contamination
by clusters associated with the bluer peak seen in the color distribution.

We estimate the contributions of the old metal-poor GCs and the
intermediate-age GCs in the following manner. 
First, we compute the total $g$ and $I$ magnitudes and \gI\ of the galaxies
using the ellipse fitting task and elapert within STSDAS.
PGC 42871 and PGC 6240 do not show evidence for their
integrated colors being significantly reddened by dust. However, PGC 6510
has redder colors and the color-index map (Fig.~\ref{colmap}) also shows strong
evidence for patchy dust distributed asymmetrically around the central region.
Since the dust distribution is asymmetric about the central region, we use
the unreddened regions to compute the reddening in the dusty regions and
correct the total magnitudes and color of PGC 6510. The corrected values
are given in Table~\ref{tbl-4}.
To estimate the gaussian LFs associated with the old metal-poor GC
population, 
we need to apply luminosity fading to the galaxies' light. Treating  
the diffuse light of the galaxies as SSPs, we estimate the luminosity-weighted
age for a solar-metallicity population from the total integrated \gI\ color index
of each galaxy and look up the expected fading in the $B$ and $V$ bands when the
galaxy ages to 14 Gyr using the GALEV SSP models (see Table~\ref{tbl-4}). A
comparison with the Maraston models is also shown in the table. The following
calculations are done using values from the GALEV models.

\begin{figure}
\plotone{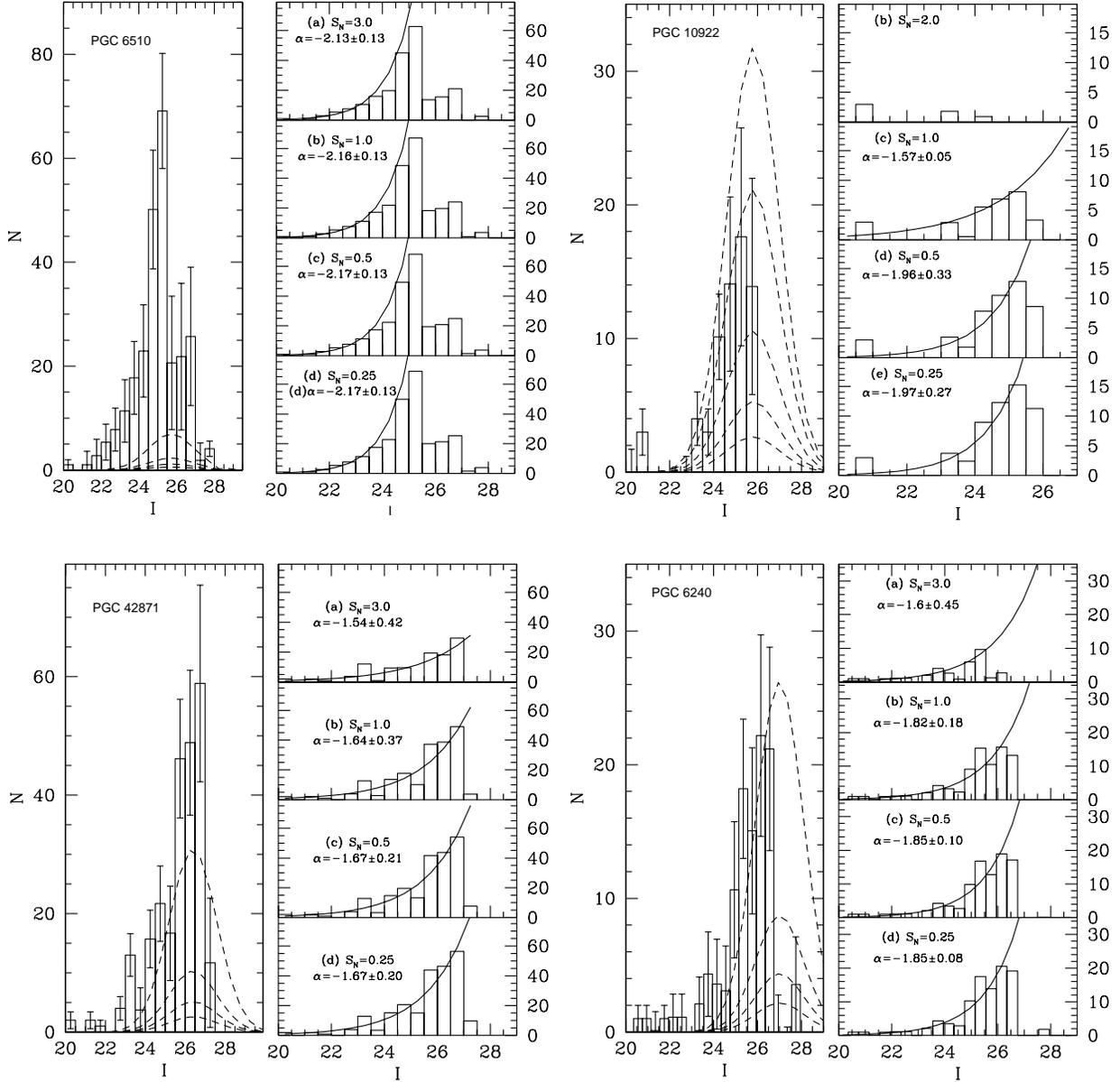}
\caption{$I$-band luminosity functions (LFs) of the GCs in the sample
  galaxies. The histogram in the left panel of each figure shows the LF of the
  galaxy corrected for background contamination and completeness. The dashed curves
  represent the estimated gaussians representing the old
  metal-poor GCs with $S_N$ values of 3.0, 1.0, 0.5, and 0.25 (from the top curve
to the bottom curve using the GALEV model),
  for PGC 6510, PGC 42871, and PGC 6240.  In the case of
  PGC 10922, the dashed curves represent the estimated gaussians
  representing the old metal-poor GCs with $S_N$ values of 3.0, 2.0, 1.0,
  0.5, and 0.25, respectively.  The histograms in the right panel for each
  galaxy show the
  residual LF obtained after subtracting the contribution of each gaussian
  from the total LF. The solid lines in the right panels are power-law fits to
  these residual cluster LFs.  The best-fit exponents $\alpha$ are
  indicated in each panel.  \label{composite}}
\end{figure}

Using the galaxy's faded $M_V$ value in the equation for the specific
frequency:
\begin{equation}
S_N = N_{\rm GC}10^{0.4(M_V + 15)}
\end{equation}
(i.e., the number of star clusters per galaxy luminosity normalized
to an absolute {\it V} magnitude of $-$15
\citep{harris81}), we calculate the number of old GCs expected for four typical
values of $S_N$ = 3.0, 1.0, 0.5, 0.25. The specific frequency is known to
increase systematically along the Hubble sequence, from 0.5 $\pm$ 0.2 for Sc
spirals to 2.6 $\pm$ 0.5 for ellipticals outside of clusters
\citep{harris}. The estimated gaussians for the old metal-poor population
were calculated using $M_I = -8.1$ and $\sigma = 1.2$ \citep{harris96,
  barmby01} for each value of $S_N$. The contributions of the gaussians for
each specific frequency are plotted in the left panel of Fig.~\ref{composite}
for each galaxy.  The residuals obtained by subtracting these estimated
gaussians from the observed LF are shown in the right panel.
We find in each case that the residuals are fit well by power laws with
exponents $\alpha$ between $-1.54$ and $-2.17$.  These exponents
are consistent with values found for young and intermediate-age cluster
systems \citep[e.g.,][]{whitmore99,whitmore02,goudfroo01b,goudfroo04,goudfroo07}. 

Figure~\ref{composite} illustrates that the observed luminosity functions
can be well modeled as composites of an old metal-poor population, as is
normally seen in early-type galaxies, and an intermediate-age population that
is more metal-rich. However, note that the relative 
contributions of these two populations to the total LF vary significantly 
from galaxy to galaxy: The clusters in PGC 6510 seem to be predominantly
intermediate-age metal-rich ones, whereas almost all clusters in PGC
10922 can be attributed to an old metal-poor population. 
This result is best seen for $S_N$ values of 1.0 and 2.0 [cases (b) and (c)
in the top right panel of PGC 10922].

\section{The Cluster System -- Host Galaxy Connection} \label{s:HostConnection}

In this section we discuss the properties of the cluster systems in the four
early-type poststarburst galaxies of our sample in terms of implications
related to the assembly of their host galaxies.

\subsection{The Nature of the Merger Progenitor Galaxies} \label{s:progenitors}

The GC color distribution of most normal early-type galaxies is bimodal,
including a ``blue'' peak with a mean color that is similar to that of
    sentence the end of Section 6.1.
metal-poor GCs in the halo of our Galaxy and a ``red'' peak with a mean color 
similar to that of the underlying diffuse galaxy light. The mean color of the
red peak has been shown to strongly correlate with the luminosity of the
parent galaxy \citep[e.g.,][]{larsen01,peng06}. According to the GALEV models, 
this peak should be seen at $g\!-\!I =$ 1.74, 1.83 and 2.02 for $Z =$ 0.2
$Z_\odot$, 0.4 $Z_\odot$, and 1.0 $Z_\odot$ respectively. However, we see no clear
indication of the presence of this ``red peak'' in any of the galaxies in our
sample.
This indicates that the progenitor galaxies lacked any significant number of
old, metal-rich GCs typically seen in normal early-type
galaxies.  
This in turn seems to point to late-type spirals as being the progenitors of
these poststarburst early-type shell galaxies since old, metal-rich clusters
are typically associated with spheroidal components of galaxies
\citep[see also][]{forbes01,goudfroo03}.  
Thus, it seems unlikely that the progenitors of these poststarburst galaxies were
ellipticals as also indicated by the presence of shells and other sharp
features indicative of former disks.

However, it is interesting to ask whether the sample
galaxies will eventually evolve into elliptical galaxies similar to
present-day normal ellipticals. A comparison of the GCs 
of present-day ellipticals with those of our sample galaxies can shed light
on this. For any such comparison, we need to take into account the changes
in the GC systems
of our sample galaxies as they evolve to older ages. One important such change
is due to cluster disruption. Star clusters are vulnerable to disruption by a variety
of processes operating on different time scales. Since the intermediate-age
clusters in our 
sample are $\sim$ 1 Gyr old, processes such as two-body relaxation, tidal
disruption, and dynamical friction all play important roles in removing stars
from the cluster \citep[e.g.,][]{fz01,vesperini03b}. 

\begin{figure}
\plotone{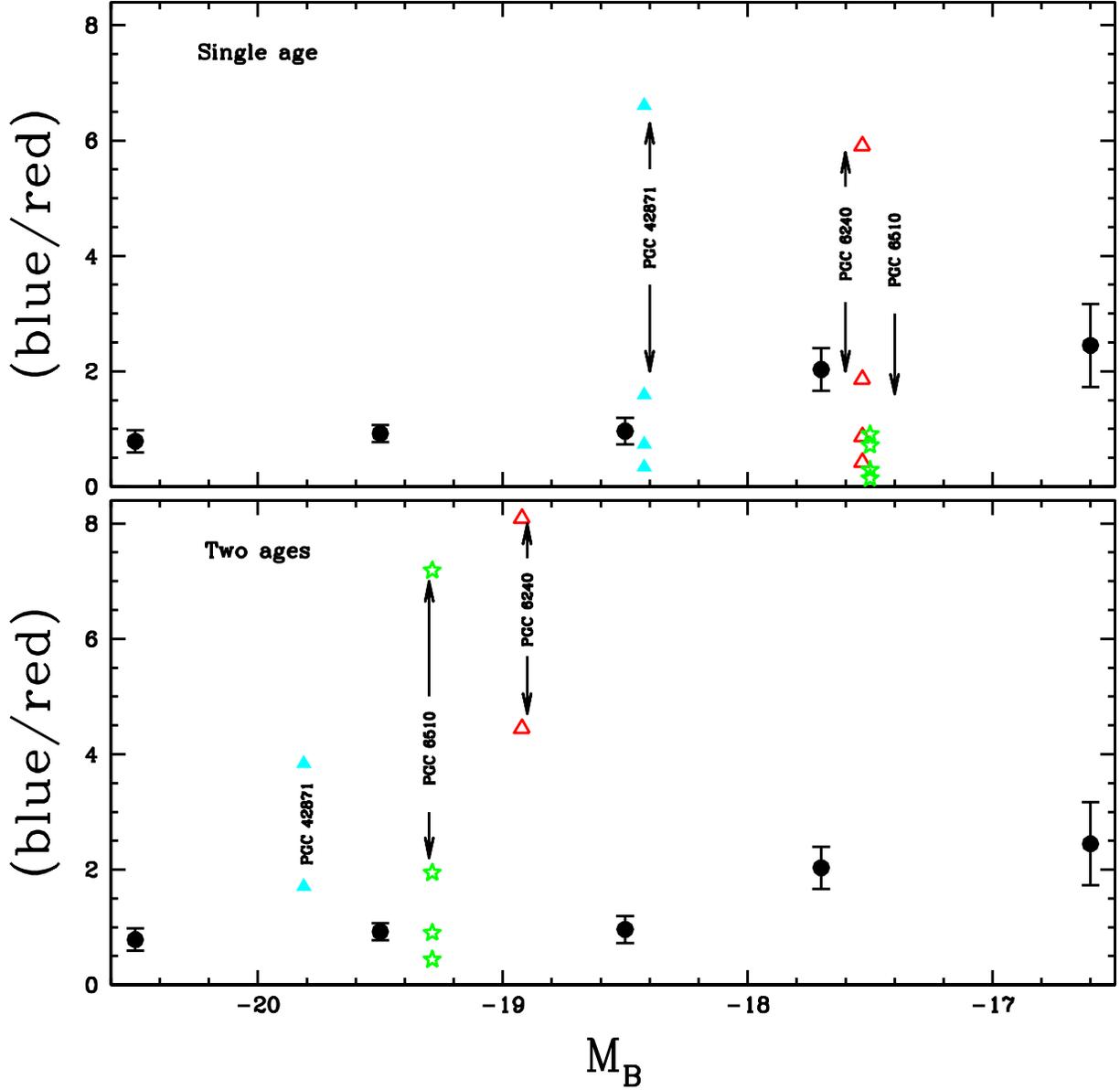}
\caption{{\it Top panel}: The number ratio of metal-poor blue GCs to
  metal-rich red GCs of the sample galaxies (using GALEV) for
  $S_N$  (= 3.0, 1.0, 0.5, and 0.25 from the top to the bottom) of the old metal-poor GCs is plotted versus galaxy $M_B$ and compared
  with values for normal early-type galaxies (black filled circles) from \citet{peng06}. The $M_B$
  values of the sample galaxies have been faded from their current
  luminosity-weighted ages to an age of 14 Gyr. PGC 10922 is not
  plotted
  because we do not detect any significant intermediate-age GC
  population in it, whereas we do so in the other galaxies. {\it Bottom panel}:
  The number ratio of metal-poor blue GCs to metal-rich red GCs (using GALEV) computed by
  considering the integrated light of the galaxies to be due to a
  superposition of old and young components. See Sect.~\ref{s:progenitors} for
  more detailed information.
\label{comparison}}
\end{figure}

In light of this, we estimate the number of intermediate-age clusters that
are expected to survive as the galaxy ages. The mass function of ``old'' GCs
is known to have a turnover at $2\times 10^5 M_\odot$. 
The number of intermediate-age clusters with masses greater than this value is
not expected to be affected by disruption to within 10\%
\citep{fz01,goudfroo07}. These relatively massive clusters 
are thus assumed to survive and become ``old'' metal-rich red GCs as the galaxy
ages.  This allows us
to calculate the number of old metal-poor GCs and the number of expected
metal-rich GCs using the results from Fig.~\ref{composite} for each of the
four values of $S_N$. We derive the 
expected number of metal-rich GCs in each galaxy by calculating the number of 
intermediate-age clusters with masses $\geq2\times 10^5 M_\odot$ given by the
power-law fit, and by applying passive evolution of the galaxies' $V$-band
luminosity to an age of 14 Gyr using the GALEV SSP models. The 
resulting ratio of metal-poor to metal-rich (i.e., blue to red) clusters 
thus obtained is compared with that in early-type galaxies of the Virgo 
cluster \citep{peng06} in the top panel of Fig.~\ref{comparison}. 
We do this for all
galaxies
except PGC 10922, which does not show evidence for the presence of any
significant intermediate-age GC population (see
Sect. 5.1 and the Appendix).  As the figure shows,
the values for PGC 6240 and PGC 42871 are consistent with those
for early-type Virgo cluster galaxies if $0.5 \la S_N \la 1$.  
Under the assumption that these early-type poststarburst galaxies will indeed
evolve to become ``normal'' early-type galaxies at old age, this finding
not only suggests that the most likely progenitors of these poststarburst galaxies
were spiral galaxies with Hubble types Sb or later
\citep[]{harris,az98,goudfroo03,chandar04}, but also provides new 
evidence based on GC system properties to support the view that mergers of
such spiral galaxies can indeed produce ``normal'' early-type galaxies at old age.   
The presence of strong tidal features in the sample galaxies (and many poststarburst
galaxies in general) also suggests disk-dominated progenitors
\citep[e.g.,][]{zabludoff96,mihos96}. 

 Note that the above analysis was performed by taking into account the
  luminosity-weighted age of the galaxies. Table~\ref{tbl-4} shows that the 
integrated colors of PGC~6240 and PGC~42871 are indeed 
consistent with those of their intermediate-age GC
populations. Hence, this assumption seems fair for those two galaxies. However,
for PGC 6510, the integrated color of the galaxy is somewhat redder than
that of the intermediate-age GC population. To evaluate 
whether this difference may account for its low values of the ratio of the number
of metal-poor to metal-rich clusters seen in the top panel of
Fig.~\ref{comparison}, we consider the alternative scenario where the diffuse
light of the galaxies is produced by a combination of an old and a
young
population (i.e., that the galaxy in question is an ``E+A''
galaxy\footnote{ Many papers now use the more general term k+a,
where  the k stands for the spectral type of an old stellar population
\citep[e.g.,][]{franx93}.}). 
In this context, we assign the young population to have an age 
as indicated by the Balmer line equivalent widths in the central spectra of
\citet[][see Appendix A]{carter88}, and to have solar metallicity. For the old
population, we estimate a metallicity by evaluating the galaxies' absolute
$K$-band magnitudes relative to that of NGC 4472, the brightest elliptical
galaxy in the Virgo cluster, and then utilizing the [Fe/H] versus galaxy $M_B$
relation of \citet[][transformed to $M_K$ using $K$ galaxy magnitudes
from 2MASS]{peng06}. Using the GALEV SSP models and the Maraston models, we then model the integrated 
\gI\ color of the galaxies (listed in Table~\ref{tbl-4}) as a linear combination
of an old component (with an age of 14 Gyr) plus a young component (with an age
determined as mentioned above). Finally, the luminosity fading of this composite
galaxy to an effective age of 14 Gyr is derived by fading only the young
component. Table~\ref{tbl-5} lists all relevant fading values for the sample galaxies. 

\begin{figure}
\plotone{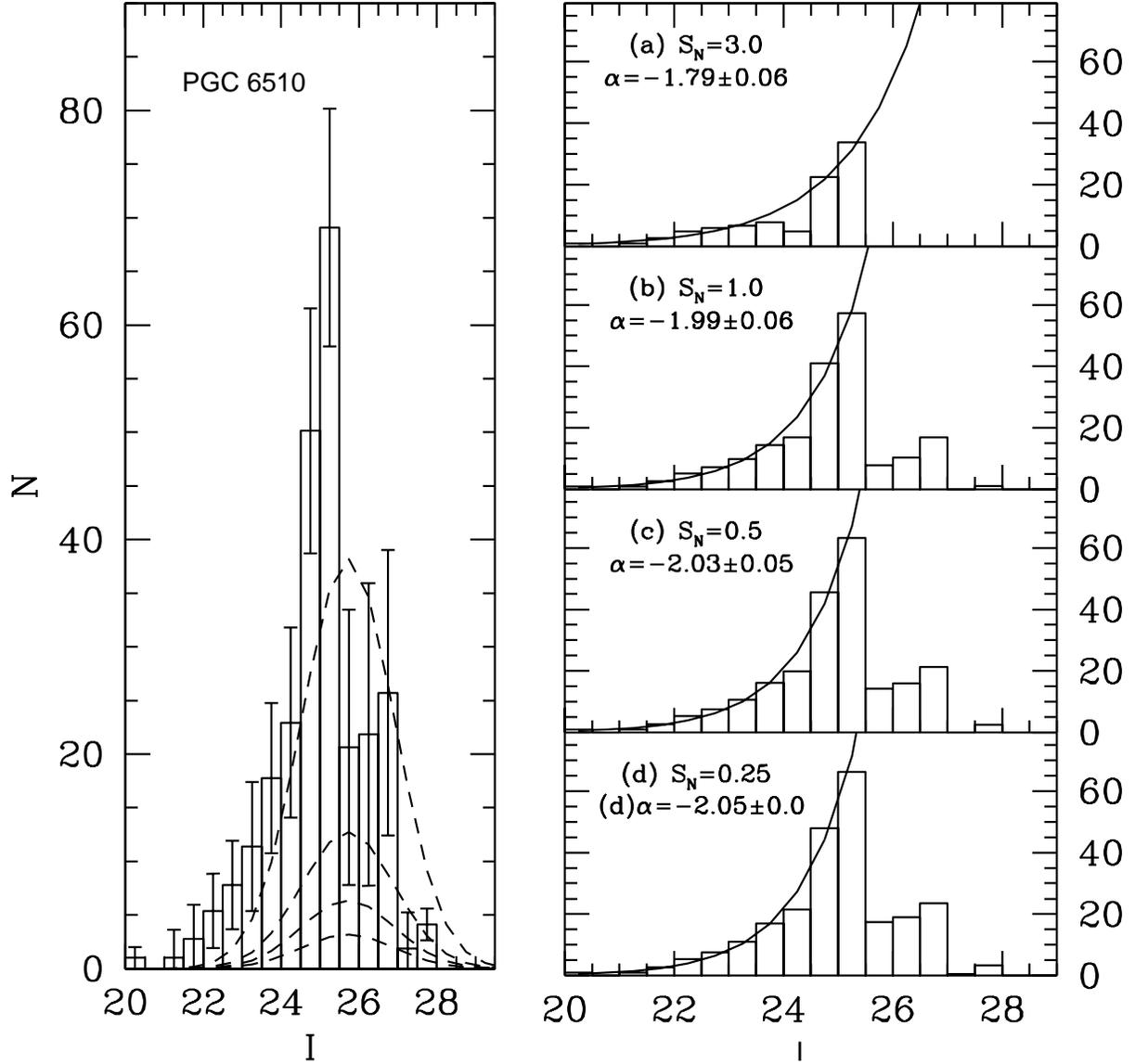}
\caption{{\it Left panel:} $I$-band LF of the GCs in PGC 6510. The histogram in the left panel
shows the LF of the galaxy corrected for background contamination and
completeness (similar to that shown in Fig.~\ref{composite}. The dashed curves represent
the estimated gaussians representing the old metal-poor GCs with $S_N$ values
of 3.0, 1.0, 0.5, and 0.25 (from the top to the bottom curve). In this case, the
diffuse light of the galaxy is modeled as a combination of an old and a young
population.(See Sect. 6.1 for details). {\it Right panel:} The histograms show the residual
LF obtained after subtracting the contributions to each gaussian from the total LF. The
solid lines are power-law fits to these residual cluster LFs. The best-fit exponents $\alpha$
are indicated in each panel. \label{p6510_compositegauss}}
\end{figure}

Using the faded magnitudes thus obtained, we repeat the procedure in Section 5.1
and model the GC luminosity function as a superposition of an old metal-poor and
an intermediate-age metal-rich population, and re-derive the predicted ratio
of the number of blue to red GCs at an age of 14 Gyr. 
For PGC 6240 and PGC 42871, this exercise results in a predicted number of
old metal-poor GCs that far exceeds the number of GCs observed, even at the
brightest GC magnitudes. This indicates that the use of luminosity-weighted
ages to model the contributions of the metal-poor and metal-rich
GCs gives more realistic results than using a combination
of ages for those two galaxies. This is consistent with the fact that the
integrated colors of those two galaxies are very similar to the mean color of
their intermediate-age GCs. 
On the other hand, for PGC 6510, we find that the
scenario involving a combination of young and old ages yields a prediction for
the number ratio of the blue to red GCs that is fully consistent with
those of normal early-type galaxies if the progenitor galaxies had $S_N
\approx 0.5$ (see bottom panel of Fig.~\ref{comparison} and 
right panel of Fig.~\ref{p6510_compositegauss}). 
The case of PGC 6510
will be discussed in greater detail in Sect.~\ref{s:strange_case} below. 
The above analysis was also done using the Maraston models. However, we do
not find significant differences between the results obtained using either
model.

\subsection{The Spatial Extent of the Poststarburst Population}

\begin{figure}
\plotone{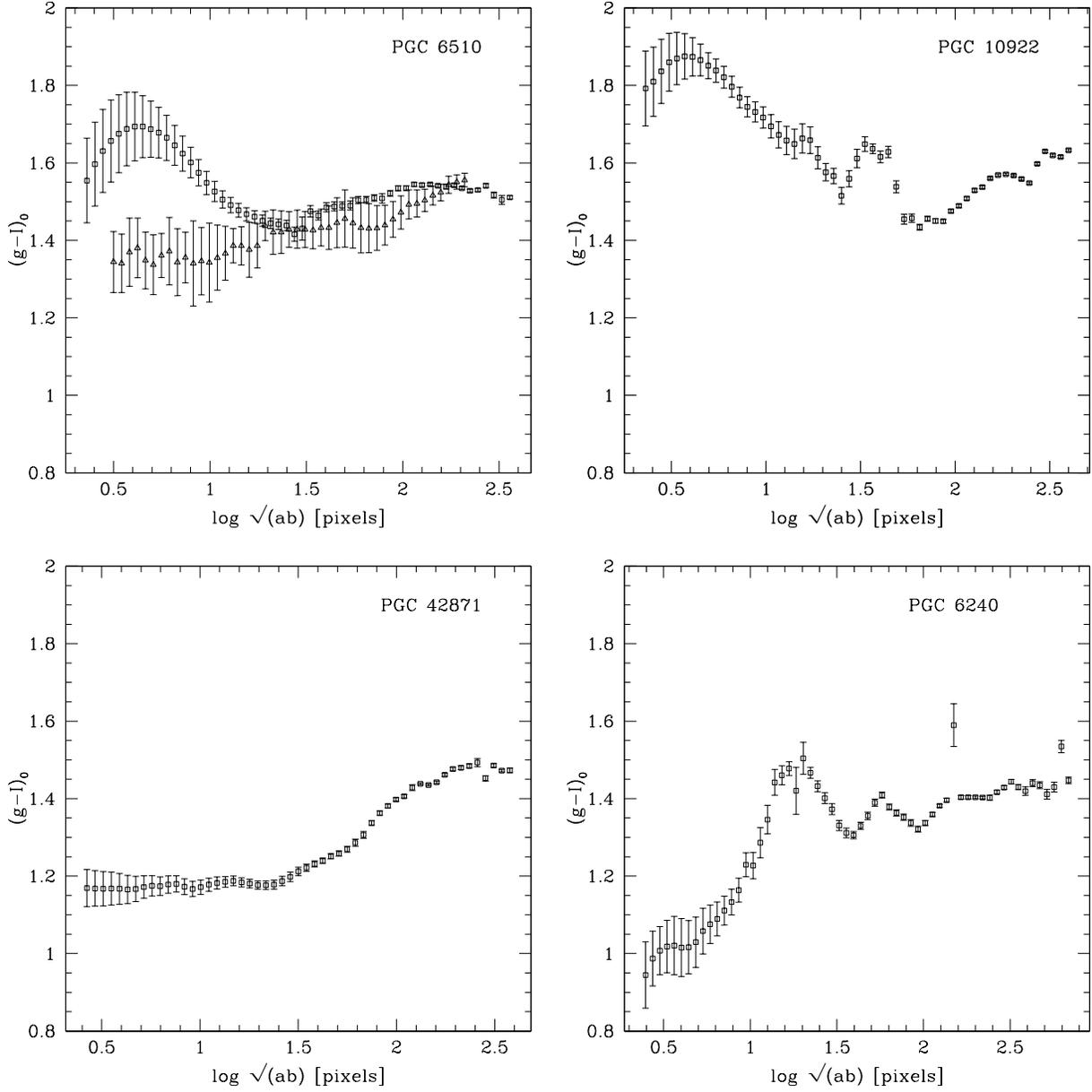}
\caption{ The radial distribution of the \gI\ color of the diffuse galaxy
  light. The central regions of PGC 6510 and PGC 10922 are redder than their
  outer regions, whereas PGC 42871 and PGC 6240 have blue central
  regions. In case of PGC 6510, the open squares represent the observed colors
and the open triangles represent the colors obtained after correction for dust. \label{radcol}}
\end{figure}

To address the spatial extent of the poststarburst population in the sample
galaxies, we use the radial color distribution of the host galaxy as a proxy.
\citet{bartholomew2001} find that poststarburst galaxies in distant clusters tend to
have slightly bluer gradients towards the center than ``normal'' early-type
galaxies.
Recent numerical simulations by \citet{bekki2005} suggest that elliptical E+A 
galaxies formed by major mergers should have positive radial color gradients
(i.e., bluer color in the inner regions).
Examining the radial \gI\ color-index profile of the underlying galaxies obtained from the
ellipse fits made in the two passbands, we find that PGC 6240 and PGC 42871 
get redder outwards, whereas PGC 6510 and PGC 10922 show redder colors in the
inner regions (Fig.~\ref{radcol}). This behavior is also seen in the
two-dimensional \gI\ color-index maps shown in Fig.~\ref{colmap}. It may well be
relevant that the two galaxies that show evidence for hosting the youngest GC
populations (i.e., PGC 6240 and PGC 42871) also show the bluest color in their
central regions. These two galaxies also have the strongest H$\delta$
equivalent widths in their nuclear spectrum among the sample galaxies
\citep[][see also Appendix A]{carter88}. 
The other two galaxies (PGC 6510 and PGC 10922) show dusty central
regions (especially PGC 10922), and it seems likely that the central reddening
in their radial color distributions can at
least partly be attributed to dust (cf.\ Fig.~\ref{colmap}).
Note that the  inner color profiles of the latter two galaxies
show a dip to bluer colors towards their very center (Fig.~\ref{radcol}),
which may well represent a signature of the poststarburst population in these
galaxies. However, Figs.~\ref{radcol} and \ref{colmap} seem to argue 
that the spatial extent of the poststarburst population in PGC 10922 and (to
a lesser extent) PGC 6510 is smaller than that in PGC 6240 and PGC 42871. We
note that this finding is reinforced by the properties of the GC systems of
these galaxies. 
We suggest that deep optical spectra of the target galaxies be
obtained to verify the spatial extent of the poststarburst population. This
would also provide quantitative information on the issue as to how well the
integrated color of the target galaxies can reliably be interpreted as a
simple stellar population, which is relevant to analyses related to the
evolution of GC specific frequencies in intermediate-age galaxies (cf.\
Sect.~\ref{s:progenitors}). 

\subsection{The remarkable GC system of PGC 6510}\label{s:strange_case}

\begin{figure}
\plotone{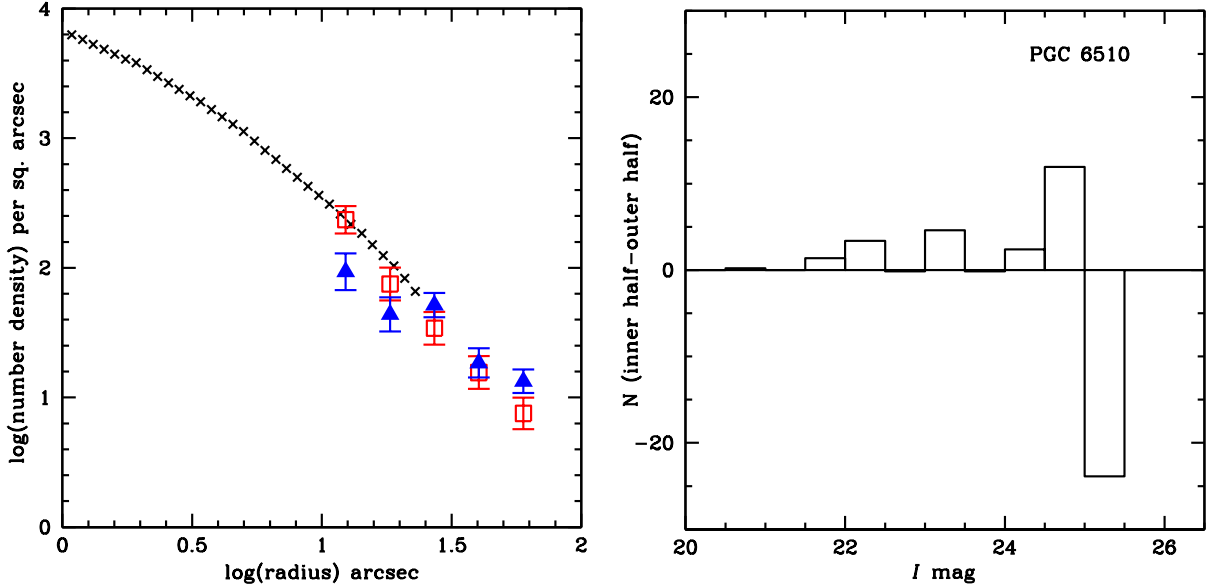}
\caption{{\it Left panel}: The surface number density of the brightest
  33$\%$ (open red squares) and the faintest 33$\%$ of the clusters
  with completeness $\ge 80$\% in the innermost radial bin (filled blue
  triangles) in PGC 6510 are plotted as a function of galactocentric
  radius. The surface brightness profile
  of PGC 6510 is also plotted on an arbitrary scale. Note that the
  profile of the bright clusters follows the galaxy surface-brightness
  profile closely. {\it Right panel}: The difference in the
  number of clusters per magnitude bin between the inner and the outer
  halves of the system of cluster candidates of PGC 6510 plotted for
  clusters
  with $I \le 25.5$ and galactocentric radius $\ge 10$ \arcsec, which
  ensures a completeness of $\ge 80$\% throughout.
  The total background- and completeness-corrected number of clusters
  in the outer half has been normalized to match that in the inner
  half. Note that the inner half of the cluster system hosts
  systematically more bright clusters than the outer half.
  \label{brt_faint}}
\end{figure}

As mentioned briefly in Sect.~\ref{s:compositeLF}, the GC system of PGC 6510 shows some
interesting properties.  
Figures~\ref{composite} and \ref{comparison} show that if the current
integrated \gI\ color of PGC~6510 is interpreted in terms of an SSP 
(i.e., a single age and metallicity), the modeling in
Sect.~5 results in the old metal-poor GCs making up
an unexpectedly small fraction of the total GC system unless the progenitor galaxies
had an unusually high specific frequency of old metal-poor GCs ($S_N \ga
3$). This result seems somewhat counterintuitive, since such high $S_N$ values
are unknown among gas-rich galaxies, whereas a significant
amount of gas was needed to trigger the starburst that led to the poststarburst
spectrum and the formation of the significant number of intermediate-age 
GCs. However, in case of the alternative scenario in which the integrated 
color of PGC~6510 is due to a superposition of a young and an old component (see
Sect.~\ref{s:progenitors} above for details), our modeling shows that
progenitor galaxies with $S_N$ values of 0.5\,--\,1 (typical for late-type
galaxies) {\it can\/} account for a more significant fraction of GCs in PGC~6510 being
of the old metal-poor kind, especially at the faint end of
the LF (compare Fig.~\ref{p6510_compositegauss} with the top left panel of
Fig.~\ref{composite}).  
To further constrain the fractions of GCs in this galaxy that are of
intermediate age versus old, 
we utilize the large size of the GC population to compare the radial surface
number density profiles of the bright vs.\ the 
faint GCs in this galaxy. 
To avoid introducing a bias due to the high background in the innermost
regions of the galaxy, we first determine the background level at which a
GC of magnitude $I$ = 25.5 is detected at $80\%$ completeness and
discard all GCs at higher
background levels for this exercise. We then
calculate the completeness-corrected surface-number densities of the
brightest $33\%$ and the faintest $33\%$ of the GCs selected this way. 
We find that the surface-number-density profile of the bright GCs 
follows the surface brightness
profile of the galaxy very closely (see Fig.~\ref{brt_faint}, left panel),
whereas the faint GCs show a more extended
radial distribution. This reinforces the idea that  
the brightest GCs are primarily intermediate-age ones since the radial profile of
intermediate-age GCs is expected to follow the surface brightness profile
of the parent galaxy \citep[e.g.,][]{schweizer96,whitmore97,goudfroo07}. We
also plot the completeness- and background-corrected LFs of the inner
half and the outer half of the GC system (within the outer radius beyond
which objects were assigned to the background) and find that the inner
half of the GC system hosts relatively more bright GCs (see
Fig.~\ref{brt_faint}, right panel). 
Both of the above results lend credence to the idea that PGC 6510 hosts a
significant number of intermediate-age GCs and that old GCs populate
the LF mainly at fainter magnitudes, as predicted by our modeling in
Sect.~\ref{s:HostConnection} for the case of a composite age structure for the
diffuse light of PGC~6510.

\section{Summary and Conclusions}

We have analyzed {\it HST/ACS\/} images of four early-type shell galaxies, for
which spectroscopy of the central region revealed a poststarburst spectrum, in
order to study the properties of their globular cluster systems. Our results
are summarized as follows.

\begin{itemize}
\item
The color distributions of the globular clusters in all four galaxies
show a broad peak centered on $g\!-\!I \approx 1.4$, while PGC 6240 and
PGC 42871 also have a significant number of GCs with $g\!-\!I \approx 1.0$.
The mean color
of the former peak is consistent with SSP model predictions for both an old
($\sim$\,14 Gyr), metal-poor ([Z/H] $\sim$\,$-1.5$) population and an
intermediate-age (1\,--\,2 Gyr) population of roughly solar metallicity. The
GCs with $g\!-\!I \approx 1.0$ are interpreted as being of age
$\sim$\,500 Myr and likely having formed in the merger.
Except for PGC 10922, the galaxies host several GCs in the
redder peak that are brighter than the maximum luminosity expected of a
single old, metal-poor ``halo'' GC of the kind commonly found around normal
galaxies.  
To test for multiple populations of overlapping color around $g\!-\!I \approx
1.4$,  we fit the observed GCLFs as composites of a Gaussian (as
seen in ``old'' GC systems) and a power law (seen in young and
intermediate-age GC systems).  By scaling the Gaussian component using
plausible values of the specific frequencies ($S_N$ values) of old metal-poor
GCs seen in present-day normal galaxies, we find the following: 

\begin{enumerate}
\item{}
We deduce the presence of a substantial population of intermediate-age
GCs in three out of four galaxies in our sample.
These GCs have ages between 1\,--\,1.5 Gyr which is comparable to
typical lifetimes of the shells, providing evidence that these GCs 
likely formed during the same merger event that formed the shells.  

\item{} 
The integrated colors of PGC 6240 and PGC 42871 are consistent with those of
the intermediate-age GCs, suggesting that the bulk of the field stars were
formed in the same star formation event that formed the intermediate-age
GCs. Interpreting the integrated light of PGC 6240 and PGC 42871 as a single-age
stellar population of solar metallicity, the ratio of the numbers of
metal-poor to metal-rich GCs are consistent with those of present-day
(giant) early-type galaxies if  
the $S_N$ values of the merger progenitor galaxies were in the range 
$0.5 \la S_N \la 1$. This range of $S_N$ values is consistent with
that of late-type spiral galaxies.  
Under the assumption that these early-type poststarburst galaxies will indeed
evolve to become ``normal'' early-type galaxies at old age, this finding
not only suggests that the most likely progenitors of the poststarburst galaxies in our
sample were late-type spiral galaxies, but also provides new 
evidence (based on GC system properties) to support the view that mergers
of such spiral galaxies can indeed produce ``normal'' early-type galaxies
at old age. 

\item{}
The integrated color of PGC 6510 is redder than that of its bright
GCs that are very likely of intermediate age, suggesting that a significant
fraction of the field stars are older than those GCs. Interpreting the
integrated light of PGC 6510 as a superposition of a young population (with
solar metallicity) and a 14-Gyr old population (with a metallicity estimated from
the galaxy's $K$-band luminosity), the ratio of the numbers of
metal-poor to metal-rich GCs is consistent with those of
present-day (giant) early-type galaxies if the $S_N$ values of the merger
progenitor galaxies were in the range $0.5 \la S_N \la 1$, just like the
situation for PGC 6240 and PGC 42871. GCs at the bright end of 
the GCLF in PGC 6510 follow the surface brightness profile of the parent
galaxy closely, supporting the notion that they are indeed intermediate-age
clusters. In contrast, GCs at the faint end of the LF (but still at 80\%
completeness) show a flatter surface number density profile, consistent with
the presence of a significant fraction of old GCs among the fainter GCs.

\item{}
We find no evidence for the presence of intermediate-age GCs in
PGC 10922, the galaxy with the smallest H$\delta$ equivalent width in
its nuclear spectrum among our sample. 
This may partly be due to the presence of a very dusty inner
region (of $\sim$\,3 kpc diameter), which prevents the detection of
any GCs there. This would imply that the star formation
activity that was responsible for the poststarburst spectrum of the galaxy was
confined to that inner dusty region. The properties of the population
of old metal-poor GCs in this galaxy are similar to those in PGC
6240 and PGC 42871.  
\end{enumerate}

\item
The color distributions of GCs in all four galaxies appear devoid
of any old metal-rich clusters, which are
generally associated with the spheroidal component of (early-type)
galaxies. This indicates that the progenitor galaxies must have lacked any
substantial bulge and were most likely of late Hubble type. 
\end{itemize}

In closing, we note that our analysis is based on $g$-- and
  $I$--band imaging and, hence, on magnitudes and one color index only.
Given the age-metallicity degeneracy among
  optical colors (e.g., Fig.~\ref{ssp}), additional spectra and/or near-infrared
icolors will clearly be needed to check the ages and metallicities
deduced for the second-generation GCs.

Early-type poststarburst galaxies are thought to represent an intermediate phase
in the formation of (at least) some elliptical galaxies from mergers. Our
results, based on properties of GC systems of such galaxies, generally
support the idea that mergers of spiral galaxies are accompanied by
the formation of clusters which evolve and form the old, metal-rich
peak of the GC color distribution as the merger remnant evolves into an
elliptical galaxy.

\appendix

\section{Notes on individual galaxies}
\begin{figure}
\plotone{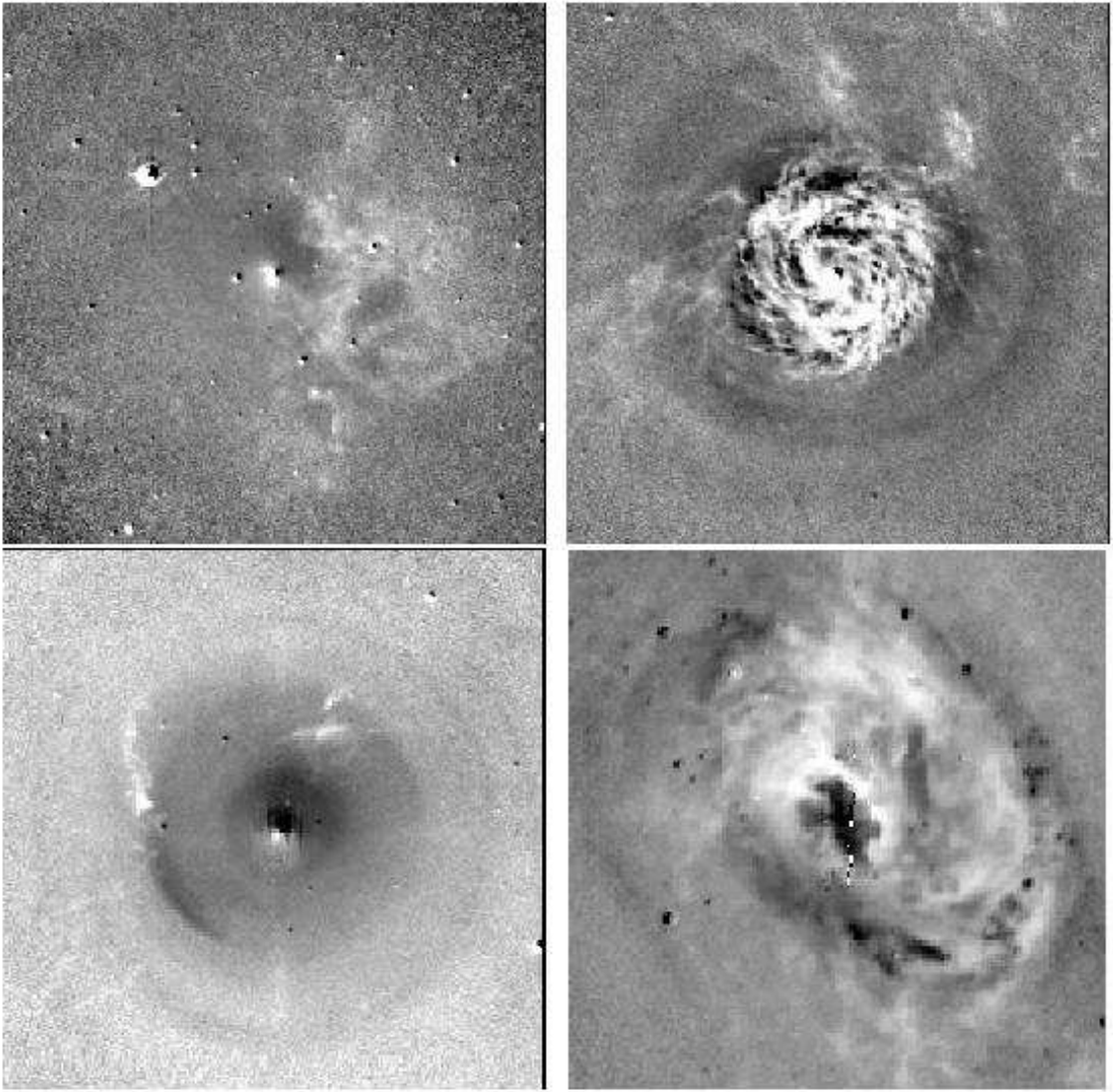}
\caption{\gI\ color-index maps of the central regions of the sample
  galaxies. Darker regions represent bluer colors and lighter regions
  represent redder colors. The images are 5 kpc on each
  side. Clockwise from top left: PGC 6510, PGC 10922, PGC 6240 and PGC
  42871. \label{colmap}}
\end{figure}

{\bf PGC 6510}:
The equivalent width (EW) of H$\delta$ in the central spectrum for this galaxy
is $5.0\pm0.8$ \AA\ \citep{carter88}. This indicates an age of $\sim$\,1 Gyr
from a comparison with the \citet{bc03} models of solar metallicity.
Using the integrated colors and the \citet{galev} models as described in Section 5.1,
the luminosity-weighted age for this galaxy is found to be 1.8 Gyr.
The GC color distribution appears unimodal. However, the LF shows an excess
of bright clusters.
This LF is modeled as a composite of an old metal-poor GC population and an
intermediate-age cluster population having the same color.
This modeling shows that most of the clusters detected in this galaxy
are of intermediate age. The \gI\ color-index map of PGC 6510 shows an
asymmetrical dust distribution (Fig.~\ref{colmap}). The properties of the
GC system of this galaxy are somewhat peculiar and do not seem to be consistent with
those of the other galaxies in the sample or with normal present-day
ellipticals, unless significant disruption of the newly formed clusters will
take place over the next $\sim$\,10 Gyr. However, modeling the diffuse
light of the galaxy as a combination of an old and a young population
gives plausible results for the properties of the GC systems.

{\bf PGC 10922}:
The H$\delta$ EW for this galaxy is $3.7\pm0.7$ \AA, which implies an age
of 1.5 Gyr from \citet{bc03}.
The luminosity-weighted age is found to be 2.2 Gyr.
The nuclear spectrum of this galaxy resembles a poststarburst spectrum but we do not
see evidence for large-scale star formation.
This galaxy has the reddest color in the central region among the sample galaxies.
This is also seen in the \gI\ map, where the central region appears surrounded
by a circular dust ring with spurs.
It seems plausible that the poststarburst population is localized near the
center and the dusty structure in this region prevents us from detecting any
clusters there.

{\bf PGC 42871}:
The H$\delta$ EW is $12.0\pm1.0$ \AA, implying an age of $\sim$\,0.3 Gyr.
The luminosity-weighted age of the galaxy is about 1 Gyr.  The color
distribution of the GCs suggests an excess of objects
at \gI\ $\approx 0.7$, which is interpreted as representing a population of
younger-age ($\sim$\,80 -- 200 Myr) clusters
based on a comparison with SSP models of solar metallicity.
The redder peak at \gI\ = 1.35 is likely made up of a composite population
of old metal-poor clusters and intermediate-age solar-metallicity clusters,
as discussed in Section 5.  Thus, we find evidence for the presence of
three separate subpopulations of GCs in this galaxy.

{\bf PGC 6240}:
The H$\delta$ EW for this galaxy is $13.6\pm 1.0$ \AA, which corresponds to
an age of $\sim$\,0.3 Gyr.  The luminosity-weighted age is found to be
about 1 Gyr. This galaxy clearly shows bimodality
in its color distribution with the bluer peak at \gI\ = 0.85 consistent with a $\sim$ 0.4 Gyr
old cluster population of solar metallicity.  The brightest shell associated
with the galaxy harbors some of the youngest clusters \citep{aparna}.  The
red clusters are modeled as a composite of
an old metal-poor population and an intermediate-age population. Hence, we
find evidence for the presence of three subpopulations of GCs in this galaxy as well.

\acknowledgements
We would like to thank an anonymous referee for helpful comments and suggestions. Support for {\it HST\/} Program GO-10227 was provided by NASA through a grant
to PG from the Space Telescope Science Institute, which is operated by the
Association of Universities for Research in Astronomy, Inc., under NASA
contract NAS5--26555.   

{\it Facilities:}  \facility{HST (ACS)}.

{}

\clearpage

\begin{deluxetable}{lllll}
\tablecaption{General properties of the sample. \label{tbl-1}}
\tablehead{
\colhead{Parameter} & \colhead{PGC 6510} & \colhead{PGC 10922} & \colhead{PGC 42871} & \colhead{PGC 6240}}
\startdata
Malin \& Carter (1983)&  MC 0148$-$836 & MC 0247$-$833 & MC 1241$-$339 & MC 0140$-$658\\
Alternative name& --&ESO 003$-$ G 013 & AM 1241$-$335& AM 0139$-$655\\
RA$^*$(J2000)&1h 46m 21.9s&2h 53m 35.9s&12h 44m 05.2s & 1h 41m 30.98s\\
Dec$^*$(J2000)& $-83\degr$ 23$\arcmin$ 59$\arcsec$& $-83\degr$ 08$\arcmin$ 32$\arcsec$& $-34\degr$ 12$\arcmin$ 07$\arcsec$& $-65\degr$ 36$\arcmin$ 55.4$\arcsec$\\
Type&E-S0&S0&S0-a & S0\\
${\it v}_{hel}$ (km s $^{-1}$) & 4652 $\pm$ 15 & 4819 $\pm$ 15 & 6074 $\pm$ 15 & 8216 $\pm$ 15\\
${\it v}_{\rm LG}$(km s $^{-1}$) & 4365& 4529& 5944 & 7936\\
Velocity disp.\ (km s $^{-1}$) & 181.9 $\pm$ 68.7 & 189.9 $\pm$  59.1 & 178.7 $\pm$ 48.8 & 249.1 $\pm$ 46.5\\
Distance (Mpc)$^{**}$ & 58.2 & 60.4 & 79.2 & 105.8\\
$m-M$& 33.82 & 33.90 & 34.49 & 35.12\\
${\it M}_{B}$&  $-$20.08& $-$20.47& $-$21.46& $-$20.57\\
${\it M}_{K_s}$& $-$23.00& $-$24.19& $-$24.65& $-$24.50\\
$A_V^{***}$&0.45&0.44&0.30 & 0.0\\
\enddata
\tablenotetext{*}{From ACS images, this work.}
\tablenotetext{**}{Using ${\it H}_{0}$ = 75 km s $^{-1}Mpc^{-1}$}
\tablenotetext{***}{\citet{burstein}}
\tablecomments{All other parameters are taken from LEDA (http://leda.univ-lyon1.fr/ or computed
using values from LEDA and ${\it H}_{0}$ = 75 km s $^{-1}Mpc^{-1}$)}
\end{deluxetable}

\begin{deluxetable}{llcl}
\tablecaption{Results of power-law fits to the GC surface density. \label{tbl-2}}
\tablehead{
\colhead{Galaxy} & \colhead{$I$-band cut-off}&\colhead{Radial range(\arcsec)} & \colhead{Power-law exponent}}
\startdata
PGC 6510&  26.0&  14 -- 137& $-1.73\pm 0.14$\\
PGC 10922& 25.8&  17 -- 178& $-1.09\pm 0.12$\\
PGC 42871& 25.0&   7 -- 115& $-1.94\pm 0.42$\\
PGC 6240&  26.6&   6 -- 108& $-1.51\pm 0.28$\\

\enddata
\end{deluxetable}

\begin{deluxetable}{ccrrr}
\tablewidth{0pt}
\footnotesize
\tablecaption{Positions and photometry of the 30 brightest GCCs in PGC 6510, PGC 10922, and PGC 42871 \label{tbl-3}}
\tablehead{
\colhead{RA (2000)}&\colhead{Dec (2000)}&
\colhead{{$I$} (mag)}&\colhead{{$g-I$} (mag)} &
\colhead{$r^{*}(\arcsec)$}}
\startdata

&& PGC 6510 &&\\
1  46   22.42 & $-$83  23  58.74 & 20.38  $\pm$ 0.01 & 1.04  $\pm$  0.01 & 1.30   \\
1  46   19.78 & $-$83  23  58.71 & 21.15  $\pm$ 0.01 & 1.87  $\pm$  0.02 & 3.22   \\
1  46   26.14 & $-$83  24  13.69 & 21.62  $\pm$ 0.01 & 1.22  $\pm$  0.02 & 16.48   \\
1  46   06.90 & $-$83  24  30.83 & 21.73  $\pm$ 0.01 & 1.18  $\pm$  0.02 & 40.34   \\
1  46   21.52 & $-$83  24  01.83 & 21.80  $\pm$ 0.01 & 1.00  $\pm$  0.03 & 2.91   \\
1  46   20.21 & $-$83  23  48.45 & 21.91  $\pm$ 0.01 & 1.30  $\pm$  0.02 & 10.62   \\
1  46   26.85 & $-$83  23  56.99 & 22.05  $\pm$ 0.01 & 1.23  $\pm$  0.02 & 9.02   \\
1  46   20.55 & $-$83  24  01.64 & 22.05  $\pm$ 0.02 & 1.45  $\pm$  0.05 & 3.31   \\
1  46   22.77 & $-$83  23  55.09 & 22.13  $\pm$ 0.02 & 1.47  $\pm$  0.03 & 4.21   \\
1  46   13.62 & $-$83  23  47.14 & 22.15  $\pm$ 0.01 & 1.65  $\pm$  0.02 & 17.97   \\
1  46   53.05 & $-$83  24  22.08 & 22.22  $\pm$ 0.01 & 0.96  $\pm$  0.02 & 58.13   \\
1  46   15.32 & $-$83  23  52.50 & 22.41  $\pm$ 0.01 & 1.68  $\pm$  0.03 & 12.52   \\
1  46   13.79 & $-$83  23  53.05 & 22.48  $\pm$ 0.01 & 1.64  $\pm$  0.03 & 14.61   \\
1  46   08.92 & $-$83  24  16.34 & 22.54  $\pm$ 0.01 & 1.43  $\pm$  0.03 & 27.72   \\
1  46   22.67 & $-$83  23  54.19 & 22.55  $\pm$ 0.02 & 1.40  $\pm$  0.04 & 4.95   \\
1  46   34.19 & $-$83  24  02.90 & 22.56  $\pm$ 0.01 & 1.19  $\pm$  0.02 & 21.69   \\
1  46   20.87 & $-$83  23  02.92 & 22.57  $\pm$ 0.01 & 1.57  $\pm$  0.03 & 55.33   \\
1  46   17.91 & $-$83  23  56.47 & 22.61  $\pm$ 0.01 & 1.35  $\pm$  0.03 & 6.83   \\
1  46   21.59 & $-$83  23  56.69 & 22.73  $\pm$ 0.03 & 1.12  $\pm$  0.06 & 2.18   \\
1  46   23.09 & $-$83  24  08.91 & 22.81  $\pm$ 0.02 & 1.61  $\pm$  0.03 & 10.19   \\
1  46   27.70 & $-$83  23  46.24 & 22.97  $\pm$ 0.02 & 1.20  $\pm$  0.03 & 16.19   \\
1  46   22.08 & $-$83  23  18.06 & 22.99  $\pm$ 0.02 & 1.03  $\pm$  0.03 & 40.35   \\
1  46   26.61 & $-$83  24  02.48 & 23.05  $\pm$ 0.02 & 1.44  $\pm$  0.04 & 9.13   \\
1  46   17.28 & $-$83  24  05.53 & 23.05  $\pm$ 0.02 & 1.19  $\pm$  0.03 & 9.94   \\
1  46   24.86 & $-$83  23  53.34 & 23.06  $\pm$ 0.02 & 1.27  $\pm$  0.04 & 7.72   \\
1  46   20.92 & $-$83  23  58.87 & 23.13  $\pm$ 0.06 & 1.73  $\pm$  0.20 & 1.27   \\
1  46   17.04 & $-$83  23  59.59 & 23.18  $\pm$ 0.02 & 1.52  $\pm$  0.04 & 7.92   \\
1  46   07.54 & $-$83  23  37.82 & 23.20  $\pm$ 0.02 & 1.45  $\pm$  0.04 & 31.83   \\
1  46   20.99 & $-$83  23  55.94 & 23.24  $\pm$ 0.03 & 1.91  $\pm$  0.11 & 3.14   \\
1  46   22.86 & $-$83  23  53.04 & 23.25  $\pm$ 0.02 & 1.72  $\pm$  0.05 & 6.13   \\

              &                & PGC 10922 &&\\
2  53   27.25 & $-$83   08  46.14 & 20.26  $\pm$  0.01 & 1.61 $\pm$  0.01 & 20.10  \\
2  52   51.47 & $-$83   09  30.32 & 20.34  $\pm$  0.01 & 1.11  $\pm$  0.01 & 97.72  \\
2  53   40.62 & $-$83   09  01.99 & 20.49  $\pm$  0.01 & 1.32  $\pm$  0.01 & 30.66  \\
2  53   14.73 & $-$83   08  19.50 & 20.82  $\pm$  0.01 & 1.68  $\pm$  0.01 & 39.62  \\
2  53   48.11 & $-$83   09  27.39 & 20.84  $\pm$  0.01 & 2.04  $\pm$  0.01 & 59.12  \\
2  54   11.37 & $-$83   08  11.53 & 20.94  $\pm$  0.01 & 1.09  $\pm$  0.01 & 67.51  \\
2  52   31.78 & $-$83   08  46.46 & 21.63  $\pm$  0.01 & 0.91  $\pm$  0.01 & 115.08  \\
2  54   28.47 & $-$83   08  24.44 & 21.74  $\pm$  0.01 & 0.74  $\pm$  0.01 & 95.09  \\
2  53   31.85 & $-$83   07  07.20 & 21.78  $\pm$  0.01 & 1.58  $\pm$  0.01 & 85.75  \\
2  53   42.42 & $-$83  10  15.81 & 21.80  $\pm$  0.01 & 1.48  $\pm$  0.01 & 103.85  \\
2  54   35.12 & $-$83   09  04.06 & 21.87  $\pm$  0.01 & 1.22  $\pm$  0.01 & 111.08  \\
2  53   46.48 & $-$83   07  37.37 & 21.99  $\pm$  0.01 & 0.76  $\pm$  0.01 & 58.67  \\
2  53   44.58 & $-$83   06  52.96 & 22.24  $\pm$  0.01 & 1.02  $\pm$  0.01 & 101.03  \\
2  52   46.70 & $-$83   08  00.83 & 22.41  $\pm$  0.01 & 1.02  $\pm$  0.01 & 93.21  \\
2  52   28.29 & $-$83   08  47.67 & 22.58  $\pm$  0.01 & 1.42  $\pm$  0.01 & 121.42  \\
2  53   21.37 & $-$83   07  18.51 & 22.71  $\pm$  0.01 & 1.76  $\pm$  0.02 & 78.43  \\
2  53   50.39 & $-$83   08  33.09 & 22.74  $\pm$  0.01 & 1.03  $\pm$  0.01 & 26.52  \\
2  52   40.81 & $-$83   09  12.94 & 22.80  $\pm$  0.01 & 1.82  $\pm$  0.02 & 105.97  \\
2  54   11.23 & $-$83   09  14.10 & 22.81  $\pm$  0.01 & 1.35  $\pm$  0.02 & 76.05  \\
2  53   35.25 & $-$83   08  58.09 & 23.18  $\pm$  0.01 & 1.75  $\pm$  0.02 & 25.41  \\
2  53   19.84 & $-$83   07  16.66 & 23.24  $\pm$  0.01 & 1.47  $\pm$  0.02 & 81.10  \\
2  53   07.95 & $-$83   07  51.23 & 23.29  $\pm$  0.01 & 1.07  $\pm$  0.02 & 64.59  \\
2  53   49.22 & $-$83   08  31.17 & 23.38  $\pm$  0.01 & 1.41  $\pm$  0.02 & 24.47  \\
2  53   35.03 & $-$83   08  21.60 & 23.62  $\pm$  0.02 & 1.62  $\pm$  0.03 & 11.13  \\
2  54   11.87 & $-$83   07  35.29 & 23.66  $\pm$  0.02 & 1.11  $\pm$  0.02 & 86.75  \\
2  53   34.44 & $-$83   08  04.21 & 23.83  $\pm$  0.02 & 1.79  $\pm$  0.04 & 28.55  \\
2  53   57.73 & $-$83   09  36.50 & 24.10  $\pm$  0.02 & 1.27  $\pm$  0.03 & 75.11  \\
2  53   22.87 & $-$83   07  11.32 & 24.12  $\pm$  0.02 & 1.45  $\pm$  0.04 & 84.51  \\
2  54   01.96 & $-$83   07  23.31 & 24.20  $\pm$  0.02 & 2.12  $\pm$  0.06 & 83.97  \\
2  53   52.48 & $-$83   08  44.55 & 24.24  $\pm$  0.02 & 2.04  $\pm$  0.06 & 32.50  \\

&& PGC 42871&&\\
12 44   03.53 & $-$34 10  29.57 & 20.25  $\pm$  0.01 & 1.77  $\pm$  0.01 & 101.29  \\
12 43   58.46 & $-$34 12  16.48 & 20.40  $\pm$  0.01 & 1.96  $\pm$  0.01 & 84.57  \\
12 44   10.70 & $-$34 11  47.39 & 20.72  $\pm$  0.01 & 1.93  $\pm$  0.01 & 70.86  \\
12 44   06.82 & $-$34 10  31.77 & 20.94  $\pm$  0.01 & 1.50  $\pm$  0.01 & 98.77  \\
12 44   02.58 & $-$34 13  11.03 & 21.27  $\pm$  0.01 & 1.96  $\pm$  0.01 & 70.68  \\
12 44   05.39 & $-$34 12  02.16 & 21.50  $\pm$  0.01 & 1.55  $\pm$  0.01 & 6.67  \\
12 44   05.29 & $-$34 12  35.41 & 21.65  $\pm$  0.01 & 1.61  $\pm$  0.01 & 26.82  \\
12 44   04.93 & $-$34 12  23.02 & 22.23  $\pm$  0.01 & 1.56  $\pm$  0.01 & 14.96  \\
12 44   06.46 & $-$34 13  16.33 & 22.27  $\pm$  0.01 & 1.34  $\pm$  0.01 & 69.37  \\
12 44   05.38 & $-$34 11  53.42 & 22.33  $\pm$  0.01 & 1.59  $\pm$  0.01 & 15.26  \\
12 44   12.35 & $-$34 12  30.94 & 22.52  $\pm$  0.01 & 1.82  $\pm$  0.02 & 90.88  \\
12 44   05.78 & $-$34 12  09.38 & 22.63  $\pm$  0.01 & 1.62  $\pm$  0.02 & 6.68  \\
12 44   05.39 & $-$34 12  53.29 & 22.66  $\pm$  0.01 & 1.74  $\pm$  0.02 & 44.72 \\
12 44   04.56 & $-$34 12  04.23 & 22.67  $\pm$  0.01 & 1.52  $\pm$  0.02 & 9.56  \\
12 44   05.32 & $-$34 12  10.99 & 22.80  $\pm$  0.07 & 0.64  $\pm$  0.05 & 2.55  \\
12 44   12.15 & $-$34 11  17.92 & 23.04  $\pm$  0.01 & 1.70  $\pm$  0.02 & 99.50  \\
12 44   03.60 & $-$34 11  01.01 & 23.14  $\pm$  0.01 & 2.11  $\pm$  0.03 & 70.62  \\
12 44   05.27 & $-$34 12  12.38 & 23.16  $\pm$  0.05 & 1.33  $\pm$  0.10 & 3.80  \\
12 44   07.06 & $-$34 11  41.88 & 23.16  $\pm$  0.02 & 1.65  $\pm$  0.02 & 34.96  \\
12 44   05.73 & $-$34 12  24.86 & 23.21  $\pm$  0.01 & 1.47  $\pm$  0.02 & 17.31  \\
12 44   05.46 & $-$34 12  23.44 & 23.24  $\pm$  0.01 & 1.41  $\pm$  0.02 & 15.07  \\
12 44   03.07 & $-$34 12  25.57 & 23.28  $\pm$  0.02 & 1.48  $\pm$  0.07 & 31.97  \\
12 44   05.86 & $-$34 12  22.47 & 23.35  $\pm$  0.02 & 1.55  $\pm$  0.02 & 15.80  \\
12 44   06.13 & $-$34 11  54.34 & 23.41  $\pm$  0.02 & 1.58  $\pm$  0.02 & 17.96  \\
12 44   04.87 & $-$34 12  15.98 & 23.43  $\pm$  0.02 & 1.59  $\pm$  0.03 & 8.76  \\
12 44   05.76 & $-$34 12  16.41 & 23.47  $\pm$  0.02 & 0.71  $\pm$  0.02 & 10.02  \\
12 44   06.17 & $-$34 10  32.79 & 23.47  $\pm$  0.01 & 1.37  $\pm$  0.02 & 96.50  \\
12 44   05.65 & $-$34 11  52.03 & 23.67  $\pm$  0.02 & 2.07  $\pm$  0.05 & 17.30  \\
12 44   04.19 & $-$34 12  20.27 & 23.78  $\pm$  0.02 & 1.73  $\pm$  0.03 & 17.59  \\
12 44   07.02 & $-$34 10  36.40 & 23.82  $\pm$  0.02 & 1.88  $\pm$  0.04 & 94.80\\
\enddata
\end{deluxetable}

\clearpage
\begin{center}
\begin{table}
\footnotesize
\caption{Fading results using a single luminosity-weighted age \label{tbl-4}}
\smallskip
\begin{tabular*}{15cm}{@{\extracolsep{\fill}}lcccccc@{}}
\hline\hline
\multicolumn{1}{c}{~} \\ [-2.ex]
\multicolumn{1}{c}{Galaxy} &{$g_0$} & {$(g-I)_0$} & {SSP Age} & {$V_{\rm est}$} & Fading in $V$ 
   & {$M_{V,\,{\rm faded}}$} \\
      & (mag)  &  (mag)      & (Gyr)     &  (mag)          &  (mag)   & (mag) \\
\multicolumn{1}{c}{(1)} &  (2)   &  (3) & (4) & (5) & (6) & (7)\\ [0.8ex] 
\hline
\multicolumn{1}{c}{~} \\ [-2.ex] 
&&&Anders \& FvA&&&\\
PGC6510&14.17&1.49&1.4&13.68&2.27&-17.86\\
PGC 42871& 13.04 & 1.30 & 1.0 & 12.61 & 2.54 & $-19.24$\\
PGC 6240 & 13.64 & 1.30 & 1.0 & 13.21 & 2.54 & $-19.45$ \\
&&&&&&\\
&&&Maraston&&&\\
PGC 6510 & 14.17 & 1.49 & 3.0  & 13.74  & 1.65
& -18.46\\
PGC 42871& 13.04 & 1.30 & 0.9 & 12.73 & 2.62 & $-19.14$ \\
PGC 6240 & 13.64 & 1.30 & 0.9 & 13.33 & 2.62 & $-19.17$

\end{tabular*}
\tablecomments{
Column~(1): Object ID. Column~(2):
Total $g$ magnitude corrected for Galactic foreground
extinction. Column~(3): Total \gI\ color corrected for Galactic foreground
extinction. Column~(4): Luminosity-weighted age
at solar metallicity. Column~(5): $V$ magnitude of the galaxy
at the age in the previous column. Column~(6): Fading in $V$ to an age of 14
Gyr. Column~(7): Faded absolute $V$ magnitude.
}
\end{table}
\end{center}

\begin{center}
\begin{table}
\footnotesize
\caption{Fading results using a combination of young and old components \label{tbl-5}}
\smallskip
\begin{tabular*}{16cm}{@{\extracolsep{\fill}}lccccc@{}}
\hline\hline
\multicolumn{1}{c}{~} \\ [-2.ex] 
\multicolumn{1}{c}{Galaxy} & Young age&\multicolumn{2}{c}{$L_{\rm frac, young}$}&\multicolumn{2}{c}{Fading (mag)}\\
&(Gyr)& $B$ & $V$ & $B$ & $V$ \\
\multicolumn{1}{c}{(1)}  &  (2)   &     (3)     & (4)       &    (5)          & (6) \\ [0.8ex] 
\hline
\multicolumn{1}{c}{~} \\ [-2.ex]
&&Anders \& FvA&&\\
PGC 6510   &   1.0  &     0.61 & 0.53 &    0.93 & 0.72\\
PGC 42871  &   0.3  &     0.58 & 0.41 &    0.92 & 0.55\\
PGC 6240   &   0.3  &     0.58 & 0.41 &    0.92 & 0.55\\
&&&&\\
&&Maraston&&\\
PGC 6510   &   1.0  &     0.46 & 0.40 &    0.61 & 0.50 \\
PGC 42871  &   0.3  &     0.47 & 0.34 &    0.69 & 0.43 \\
PGC 6240   &   0.3  &     0.47 & 0.34 &    0.69 & 0.43

\end{tabular*}
\tablecomments{
Column~(1): Object ID. Column~(2):
Age of young component as derived from Balmer line strengths in central spectrum (cf.\
Appendix A). Column~(3): Current luminosity fraction of young component in the $B$
filter. Column~(4): Same as column (3), but in the $V$ filter. Column~(5): Fading of
composite population (due to young component) in $B$ to an age of 14 Gyr. 
Column~(6): Same as column (5), but in $V$.
}
\end{table}
\end{center}

\end{document}